# The Price of COVID-19 Risk In A Public University

Duha T. Altindag

Samuel Cole

R. Alan Seals, Jr.*

March 2022

## Abstract

We study the allocation of and compensation for occupational COVID-19 risk at Auburn University, a large public university in the U.S. In Spring 2021, approximately half of the face-to-face classes had enrollments above the legal capacity allowed by a public health order, which followed CDC's social distancing guidelines. We find lower-ranked graduate student teaching assistants and adjunct instructors were systematically recruited to deliver riskier classes. Using an IV strategy in which teaching risk is shifted by classroom features (geometry and furniture), we show instructors who taught at least one risky class earned $7,400 more than those who did not.

Keywords: COVID-19, Occupational Risk, Instrumental Variables, Behavior of Elites
JEL Codes: J31, J45, I23

* Authors are all affiliated with the Department of Economics at Auburn University. Altindag: altindag@auburn.edu; Cole: sdc0038@auburn.edu; Seals: alan.seals@auburn.edu. We thank Joseph Compton, Connor Fogal, Meagan Jones, Audrey Millhouse, Nic Recasens, and Walt Swan for excellent research assistance. We are indebted to Randy Beard, Beth Davis-Sramek, David Laband, Ian Schmutte, Michael Stern, Traci Witte and seminar participants in various universities for their helpful comments. Several Auburn University administrators provided critical institutional support and expertise: General Ronald Burgess, Ricky Causey, Jamie Hammer, Bill Hardgrave, Daniel King, and Simon Yendle. Any errors in this paper are our own.

# 1. Introduction

Plans to restore normalcy to college campuses shuttered by the COVID-19 pandemic divided faculty, administrators, students, parents, and civic and business leaders during the 2020-2021 academic year. To balance stakeholders' considerations, the U.S. Center for Disease Control (CDC) created reopening guidelines for university operations, outlining key policy considerations for testing, social distancing, and contact tracing.[1] State governments quickly responded with their own COVID-19 rules and orders for university reopenings.

We study the reopening of Auburn University, a large public research university in Alabama. On May 21, 2020, Alabama Governor Kay Ivey issued a public health order with specific instructions for the resumption of on-campus activities at the state's universities.[2] Concerning social distancing, the order stated: "Schools shall take reasonable steps, where practicable, to maintain six feet of separation between persons of different households." During the Summer of 2020, the Auburn University Architect's Office began to conduct Classroom Capacity Analyses (CCAs), presumably to comply with the Governor's social distancing order.

Auburn University resumed in-person classes in Fall 2020 on a voluntary basis and, following the low participation of instructors, mandated in-person classes for instructional faculty in the Spring 2021 semester.[3] However, the CCAs were not made available to the university community until December 18, 2020, after the Spring 2021 schedule was set.[4] After the faculty scrutinized the plans for mandated face-to-face instruction, administrators acknowledged, in recorded public forums, that the university did not have a social distancing policy for its classrooms. Instead, the central administration relied on a 50-percent enrollment limit of the normal seating capacity in classrooms (50% Rule).[5]

---

[1] With respect to on-campus activities and social distancing, the CDC categorizes different strategies from "least" (no in-person learning) to "highest" (e.g., in-person classes and campus gatherings with no social distancing and/or mask requirements) risk. Link to the relevant CDC web page.
[2] See Governor Ivey's May 21st, 2020 "Safer at Home Order." Specifically, point 13 relates to universities. Also, see Governor Ivey's January 21, 2021 extension of the order until March 5, 2021.
[3] Concerns regarding Auburn University's closure during the Spring 2020 semester and subsequent reopening in the Fall 2021 semester were documented by local news outlets. Discussions centered around the survival of local businesses, University finances, and the safety of students and faculty. We provide greater details in Appendix A.
[4] To the best of our knowledge, upper administration acknowledged the CCAs on January 27, 2021, in weekly COVID-19 update email from the university, after the semester started. Thus, some instructors were unaware of the risk.
[5] No other policy concerning social distancing in classrooms was initiated. Moreover, instructors were explicitly informed, at a public forum, they could not enforce social distancing in their classes. See Appendix A.

We compared the enrollment limits from the CCAs with reported student enrollments in face-to-face (F2F) classes.[6] About half of the F2F courses in Spring 2021 were delivered in what we refer to as "*risky*" classrooms—where the enrollment exceeded the maximum number of students that can be seated in the room while allowing six feet between everyone. Also, approximately one out of four Spring 2021 F2F classes violated the university's 50% Rule.

Utilizing the granular nature of our risk measure, we investigate the association between instructor characteristics and the probability of teaching a risky course. Graduate student teaching assistants (GTAs) and non-tenure-track adjunct instructors were more likely to teach risky classes in Spring 2021 than the tenure-track professors and administrators teaching in the same department, holding constant personal attributes (including age) and course characteristics. The magnitude of this effect is equivalent to about one-third of the mean of the outcome. We also show that the relationship between instructor attributes and the (hypothetical) risk did not exist in Spring 2020, implying lower-ranked faculty do not routinely teach more crowded classes than their higher-ranked counterparts. These jointly suggest a systematic effort in Spring 2021 to recruit certain types of instructors to deliver risky courses.

Were instructors compensated for the classroom risk? Rosen's (1986) equalizing wage differentials model provides a natural theoretical starting point to understanding the instructional wage structure at Auburn. However, that model implies a market with perfect information. In contrast, the Spring 2021 market for risk at Auburn University was much more complicated.[7] Department chairs generally perform the teaching assignments of the instructional staff. These department chairs are local monopsonists, as instructional labor is usually not substitutable across academic departments, and Auburn University is the largest academic employer within 100 miles. In addition, instructors' knowledge of classroom risk was imperfect when courses were scheduled due to the late announcement of the CCAs. The central administration also lacked information about individual faculty members' risk tolerances and attitudes. In response to these modeling constraints, we outline an alternative framework in which department chairs determine instructors' wages and classroom risk to achieve the central administration's objective to minimize the total cost, subject to delivering at least a certain amount of in-person classes. Our empirical estimate of the price of COVID-19 risk below is best viewed as the *central administration's valuation of risk*.

---

[6] CCAs are discussed in the Data section, with examples in Appendix A Figures A2 and A3.
[7] In an earlier version of this paper, we organized our thoughts around Rosen's compensating differentials model while acknowledging the existence of imperfect information. However, subsequent to discussions with Randy Beard, emeritus professor of economics at Auburn University, and others, we reconsidered that position.

Using data from the university's payroll records and classroom risk assignments of instructors, we estimate the wage premium for risk. We employ an IV strategy to guard against potential selection bias associated with information asymmetries about risk and unobserved heterogeneity across instructors. Our instrument, *Dispersible Class*, indicates whether the classroom characteristics (layout) or furniture (attached to the ground or movable) allow students to spread out and away from each other when attending the lectures. Importantly, it has a mechanical relationship with our measure of risk. Controlling for its square footage, the COVID-19 enrollment capacity of a classroom is higher in dispersible classes.[8] As a result, courses held in rooms where students can spread out are expected to be less risky. Indeed, we show a solid first-stage relationship between our instrument and the level of risk. Our identification strategy relies on the assumption that the IV is related to an instructor's wages only through the risk of the class she/he teaches. That assumption is likely to hold in our application because classroom layouts were determined long before the teaching schedules. Further, we show that instructors' or her/his class's observable attributes are not correlated with whether the assigned classroom is dispersible, suggesting that our instrument is as-good-as-random.

Results of 2SLS regressions show that instructors who teach at least one risky class earn about 20% more than their observationally equivalent counterparts who do not deliver any risky courses. The risk premium is approximately $7,400 in a four-month semester. However, these results do not imply that every instructor who taught a risky class was compensated for the risk. In a placebo test, we find that instructors who taught hypothetically risky courses in the pre-pandemic Spring 2020 semester were not paid more than those who taught safe courses, suggesting that the risk premium existed only in the Spring 2021 semester and not before.

Below, we first present critical institutional details, followed by a description of our data and empirical analysis. Next, we demonstrate that Auburn University saved approximately $8 million in labor costs by ignoring CDC's social distancing guidelines. The last section offers a summary and discusses how our paper relates to the literature.

## 2. Background and Conceptual Framework

---

[8] The Director of Facilities at Auburn University also confirmed this relationship in a meeting. See Appendix A.

*Institutional Background*

Auburn University, one of the two large public universities in Alabama, is a (Carnegie R1) research university established in 1856. About 30,000 students attend the university and constitute a significant portion of the city of Auburn's population. Like other college towns, the Spring 2020 shutdown was a substantial blow to the university's budget and the local economy.[9]

The economic costs of the pandemic and pressure from parents and students who questioned the effectiveness of remote instruction appear to have incentivized the decision-makers at Auburn University to increase the number of in-person classes in the 2020-2021 academic year.[10] During the Summer of 2020, university officials began preparations to resume in-person instruction.[11] Starting in mid-June and throughout the Fall 2020 semester, the university conducted surveys of its classrooms to determine the maximum number of students who can be seated in the classrooms while maintaining six feet of social distance. The timeline of these studies, according to the date stamps on the documents, is depicted in Appendix Figure A4.[12]

In the Fall 2020 semester, the university initiated its 50% Rule, which allowed classrooms to contain up to half of their usual enrollment limits.[13] For the Fall 2020 semester, the university administration asked instructors to deliver their courses in the modality they thought best aligned with their objectives. Only 19% of classes were taught in person in that semester. Despite the low in-person activity, within a few weeks of the semester, Auburn University's main campus experienced many COVID-19 cases, 1,074, the fourth largest in the country.[14]

In the Spring 2021 semester, the university required all instructors to deliver their classes in pre-pandemic modalities (typically F2F) unless they had an exceptional medical excuse or a compelling pedagogical counter reason. As a result of this push for in-person classes, 71% of the courses were taught in person in Spring 2021. It is noteworthy that the full set of CCAs was completed by December 18, 2020, before the Spring 2021 semester began in January, albeit after teaching schedules had been finalized. Thus, the university administration knew that many classes contained more students than their rooms' lawful capacities, according to its own measurements. Despite this fact, the administration did not change the 50% Rule for the Spring 2021 semester.

---

[9] Appendix A Figure A1 shows that Auburn's retail and recreation businesses suffered the greatest among all Alabama college towns. Appendix A provides details about the impact of the Spring 2020 shutdown.
[10] See for example Altindag, Filiz, and Tekin (2021), Bird, Castleman, and Lohner (2020), and Kofoed et al. (2020) for studies that investigate the effectiveness of online classes.
[11] These considerations were discussed in the University Senate's meeting on June 16, 2020.
[12] See Appendix A for a more complete explanation of the evolution of the Auburn administration's position on social distancing. Although the university administration discussed conducting CCAs in June 2020, they were not completed before the Fall 2020 semester started. There was a hiatus in their implementation between August 11 and November 18. Some argue that CCAs were restarted on November 18 because of a call for vote of no confidence in the provost on November 10. Despite our several inquiries, the administration did not explain the peculiar timing of the studies.
[13] By contrast, the capacity of the open-air Jordan-Hare Stadium was restricted to 20% in the 2020 football season.
[14] This is based on an analysis that appeared in the popular media. Also see University's historical data on infections.

After the management of the reopening was scrutinized at several Auburn University Senate meetings, a motion for "No Confidence" in the Provost was made on November 10, 2020. The debate around the motion was covered extensively in the press and divided the university community. The administration responded with a series of Town Hall meetings. In December 2020, concerned instructors were encouraged to divide the registered students in their risky classes into groups with rotating attendance or deliver their lectures in other safer modalities.[15] This effort to assuage faculty concerns also increased the instructors' workload and introduced potential problems for instructors teaching a class differently from its advertised modality.[16] In public forums, the university administration defended its policies by underlining that there were no documented cases of COVID-19 transmission in classrooms, *ex-post*.

*Conceptual Framework*

The market for instructional services at Auburn University is not competitive. Labor is generally not substitutable across academic departments, and, as a result, the department chairs are effectively local monopsonists. Furthermore, the instructors were unaware of elevated risks in their classrooms in the Spring 2021 semester. A lack of bargaining power for instructors, coupled with incomplete risk information, does not lend itself to a model of bargaining over the risk premium.[17]

The primary decision-maker in the Auburn University setting is the central administration which faced significant pressures to increase F2F course offerings. We assume a target level of in-person courses emerged due to these pressures. The administration's problem involved minimizing the total cost subject to providing this target level of F2F classes. However, the administration lacked the requisite knowledge of course characteristics and instructors' risk tolerances and expertise. Thus, the administration recruited the department chairs to help solve its problem.

---

[15] However, these directions to instructors directly contradict an October 8th, 2020, letter, addressed to the students and signed by the Provost, stating that most classes would be in-person and that the advertised teaching modalities would let students "know when you register how each course section will be delivered".
[16] For example, students who specifically signed up for a F2F class, perhaps because they learn the best in F2F, may suffer when their teachers deliver lectures online. Such students may file grievances or provide the instructor with an unfavorable course evaluation at the semester end. In addition, if practices such as delivering a class in a modality different from what was advertised are considered fraud, then the instructor and/or the university could be liable.
[17] Loertscher and Marx (2022) discuss the challenges of bargaining problems with incomplete information.

Instructor $i$'s total pay is $\bar{w} + w_i$. The department chair cannot alter $\bar{w}$ (the standard/fixed wage for the instructor) but she/he can choose a wage premium, $w_i \geq 0$. In addition to the wage bill, the university's costs include the expected cost associated with "trouble" for the university, i.e., lawsuits, disparaging newspaper articles, and other public fallout caused by the instructor. This expected trouble cost, $f(x_i, w_i, r_i)$, depends on the instructor's attributes ($x_i$), classroom risk assignment ($r_i$), and $w_i$. One would expect $f(\cdot)$ to be larger for tenured faculty, as they have better outside options, higher bargaining power, and more significant political clout within the university. Other things same, a higher $r_i$ increases $f(\cdot)$, while a larger $w_i$ reduces it.

Ignoring the costs associated with the physical plants, the administration wishes to minimize the total cost, $C = \sum(\bar{w} + w_i + f(x_i, w_i, r_i))$. Hence, the chairs choose the risk allocations and wage premiums of instructors, given that $\frac{\partial f(\cdot)}{\partial w_i} = -\$1$ at the cost-minimizing optimum. From this process, the optimal wage premia $w_i^*$ and risk allocations $r_i^*$ are determined. Some instructors may be assigned risky classes and receive high wage premia, while others receive little or none. Note that $w_i^*$ represents the sufficient incentives for instructors to accept the risk and, thus, must be at least as large as the instructors' reservation wage. More importantly, $w_i^*$ represents the optimal expenditure by the university to minimize the costs of "trouble" and thus are analogous to expenditures on self-protection in the risk and insurance sense (Ehrlich and Becker 1972). In our empirical analysis, we estimate the relationship between $w_i^*$ and $r_i^*$.

*Demand for and Supply of Instructors in Spring 2021*

Besides the riskiness of the classes, other important factors may influence the wages of instructors in the Spring 2021 semester. One of these determinants could be the demand for instructors at the university. For example, due to the university's 50% rule, which requires classes not to exceed half of their regular seating capacity, one would expect the university to offer more sections of courses. As a result, more instructors would be needed to teach these courses, and the growth in demand for instructors would increase prices in this labor market.

We investigated this possibility and found counterevidence. Specifically, our analysis shows that the number of classes offered in Spring 2021 (8,013) is not higher than in Spring 2020 (8,217). The average capacity of (the number of students allowed in) classes fell only negligibly from 23.69 in Spring 2020 to 23.47 in Spring 2021. Similarly, the average number of students who registered in a class also did not change significantly from 2020 to 2021 (17.38 to 17.59). This is probably because the 50% rule was not implemented universally at the university.[18] We find that approximately *25 percent* of F2F lecture-based classes in Spring 2021 violated the 50% rule.[19] We also find that the composition of course offerings is similar when comparing those offered in Spring 2020 and 2021.[20] All of these suggest *no significant change in the demand for instructors in Spring 2021* compared to previous semesters.

We further studied whether there was a change in the supply side in Spring 2021 that may have affected wages. For example, some instructors may have opted out of teaching in Spring 2021 due to the fear of COVID-19 risk. Under this scenario, wages would rise because of instructor shortage. However, we do not find evidence supporting this either. For example, the number of unique instructors who taught at least one class in Spring 2021 was 1,968, virtually identical to that in Spring 2020 (1,986). The attributes of the teaching body in 2021 versus 2020 are the same as well. For example, among those who taught at Auburn University in Spring 2020 (2021), 41.3% were female (42.4%), 21.9% were non-white (21.7%), 58.9% were tenured or on the tenure track (61.4%). The average age was 43.5 (42.8). We conclude that *there was no significant difference in the composition or the size of the instructors* in 2021 versus the pre-pandemic semesters.

## 3. Data

Our data set is compiled from a number of publicly available sources. Course information, such as the course's code or department (e.g., ECON), level (e.g., 101), instructor of record, classroom (e.g., Miller Hall room 207), and mode of delivery (modality), is obtained from the Dynamic Scheduler on Auburn University's website.[21] We also gathered the capacity of the class

---

[18] On December 3rd, 2020, the Provost sent an email to the academic deans at Auburn stating, "…*have received some reports this week of potential classroom-related issues for the spring* [2021 semester]. *First, in some instances, department chairs/heads (or someone from the colleges/schools) are increasing enrollment capacities in sections beyond the 50% capacity without discussing with the faculty members.*" We obtained this correspondence through an open records request, which is available upon request.

[19] We computed the normal classroom capacities using the 2019-2020 academic year enrollment data. The half of these normal capacities determine the maximum number of students who can attend the class while complying with university's 50% rule. We found over 500 F2F classes (out of 1,512) in Spring 2021 violated the 50% rule.

[20] In Spring 2020 and 2021, 3,054 and 3,160 unique courses (e.g., ECON 101) were offered, respectively. 631 courses are only offered in Spring 2021 and not in Spring 2020. 525 courses are only offered in Spring 2020 but not in Spring 2021. 2,529 courses are offered in both semesters.

[21] The modalities were F2F flexible, F2F required, blended flexible, blended required, and online/distance learning. F2F required courses are in traditional class format, whereby student attendance is mandatory. For F2F flexible

(the maximum number of students that could enroll as permitted by the university) and the actual number of students registered for the course from this source.

We define a class as a section of a course. In Spring 2021, Auburn University scheduled 8,013 classes, 996 of which were not assigned a modality. These are usually independent study, thesis, or dissertation courses. As shown in Table 1, the vast majority of the remaining 7,017 classes were advertised as F2F (71%) in Spring 2021. Only 24% and 5% of the classes were taught in online and mixed modalities, respectively.

The second source of our data is the Office of the University Architect, which conducted the [Classroom Capacity Analyses](#) (CCAs). These surveys examine each classroom's geometric and spatial characteristics and determine the number of students that can fit in them and maintain the CDC prescribed six-foot social distance from others. We obtained all 395 CCAs on the university's website. Two examples of CCAs are provided in Appendix Figures A2 and A3 (rooms 207 and 230 in the Miller Hall – home to the Department of Economics at Auburn University).

We compare the CDC-recommended classroom capacities with the enrollment in classes and categorize classes according to whether the number of enrolled students surpasses the safe capacity.[22] *Safe Classes* are those in which the number of enrolled students is less than the CDC-recommended capacity of the classroom where they are taught. In the *Risky Classes*, enrolled students are greater than the CDC capacity. We grouped classes for which registered students are at least twice as many as the safe capacity as the *Very Risky Classes*.

A critic may argue that teaching a *Risky* or a *Very Risky* class does not necessarily mean the instructor or the students are are at an immediate and high health risk. We agree with this viewpoint. Instead, we think of our measure of risk as *an undesirable job attribute*. We postulate that teaching a safe class is preferred over teaching a risky class, other things equal. In addition to the potential COVID-19 complications for the students and those with whom they live, teaching in a packed room may increase an instructor's workload. For instance, they may feel compelled to take safety precautions or deliver their classes in multiple modalities, potentially different from the announced modality. If students who specifically signed up for a F2F class suffer when their

---

courses, student attendance is expected by the instructor, but the instructor may grant leniency if the student is not able to attend. Further information regarding Spring 2021 modalities can be found at this [web page](#).

[22] We could not categorize all of the classes into these groups. This is because for some classrooms, no CCA was performed, and for some other classes there was no assigned room, presumably, because they do not take place in a classroom, such as independent study, thesis, or research and dissertation credit hours. Typically, very few students are registered to these classes.

teachers deliver their courses in an online or mixed modality, they may file grievances or provide instructors with an unfavorable course evaluation at the end of the semester. In addition, delivering a class in a modality different than its advertised modality may be considered fraud.[23]

Table 2 presents the distribution of the riskiness of the classes at Auburn University by their modality in Spring 2021. In this table, we analyze only a subset of the courses for which we could compute the risk.[24] Statistics in this table reveal that *in about half of the F2F classes, a six-foot distance between students could not be maintained*. In a staggering 22% of classes, enrolled students are at least twice as many as the safe capacity. Online courses pose no risk to students or the instructors, as they allow remote class participation.

We augmented our class data with several instructor attributes, such as their race, sex, and age. The sources of these data include the instructors' online profiles on the university's web page and other web pages such as LinkedIn or ResearchGate.[25] We obtained titles and salary information for the instructors in our sample from the Database of the Auburn University Employee Salaries, the People Finder on the University website, and the [payroll records](.) of the university provided by the Controller's Office. Payroll records allowed us to compute each instructor's work experience at the university (the number of years between the date of the instructor's earliest paycheck from the University and January 1, 2021). Title, salary, and experience information was obtained for about 99% of all the instructors.[26]

In Table 3, we present the descriptions and the summary statistics of the attributes of classes offered in Spring 2021. The unit of observation is a class. Only those we could categorize as *Safe*, *Risky*, or *Very Risky* are considered for this analysis. On average, 26 students are registered in a class that takes place in a classroom with a safe capacity of 21. The majority of the courses (69%) conform to the CDC guideline of six-foot social distancing. The remaining classes are categorized as *Risky*. About 60% of classes are taught in-person (*F2F*), and a significant portion of them are delivered *Online*. About two-thirds of the classes are undergraduate classes. We present the number of students and the share of risky courses by lecture-type (lecture, lab, practicum, and so on) and by the departments (Economics, English, and so on) in Appendix A Tables A1 and A2.

---

[23] According to the Alabama Code (6-5-101) "Misrepresentations of a material fact made willfully to deceive, or recklessly without knowledge, and acted on by the opposite party, or if made by mistake and innocently and acted on by the opposite party, constitute legal fraud."

[24] The risk in some classes could not be identified primarily because they were not assigned a room. These courses are typically dissertation or research hours or independent study courses.

[25] From their online profiles, we obtained each instructor's photo. Three separate reviewers (undergraduate students) classify the race (white or non-white) and sex (male or female) and estimate their age from the anonymized images. We categorize the race and sex of an instructor by the majority opinion and use the mean of the three age estimates in our analysis. There was unanimous agreement on race and sex for 93% and 99% of the time. The correlation between the age estimates is 0.75. About 8% of the instructors' photos could not be located. We relied on their first and last names, the [Social Security birth records,](.) and the [most common surnames from the 2010 US Census](.) for these instructors' attributes. For the few instructors with uncommon names, we utilized [other sources.](.) Ultimately, we imputed the race, sex, and age of 100%, 99.4%, and 91.8% of the instructors, respectively.

[26] In all of the regressions that follow we control for instances of missing instructor attributes, including missing values for sex, age, title, and experience, using separate indicator variables.

Table 4 shows the summary statistics of instructor attributes. The unit of observation in this table is an instructor. Approximately 40% of the 1,501 instructors who taught in Spring 2021 are *Female*. Slightly less than one in four of them is *Non-White*. About 60% of the instructors are tenure-track, *Full*, *Associate*, and *Assistant*, professors. The remaining are *Lecturers* (who are holding tenure-track teaching positions), *Adjunct Instructors* (who are holding non-tenure-track teaching positions), graduate student teaching assistants (*GTA*s), *staff* (who separately work elsewhere in the university), and *Administrator*s (who also have administrative posts, such as department chairs, deans of the colleges, and those in the university's upper administration). Typically, an instructor teaches four classes in Spring 2021. As part of their compensation, instructors earned $9,700 monthly. The typical instructor has been employed at the university for more than five years and is in her/his early 40s.

In our empirical analyses below, where we estimate hedonic wage regressions, we use the attributes of the "riskiest class" taught by the instructor. To identify their riskiest class, we computed the ratio of registered students to the CDC-prescribed safe classroom capacity for all the courses taught by an instructor. The one with the highest ratio is the instructor's riskiest class.[27] In the bottom panel of Table 4, we present the statistics about the instructors' riskiest classes.

## 4. Empirical Analysis
### 4.1. Who Taught a Risky Class in Spring 2021?

To study the relationship between the probability of teaching a *Risky* class and the individual attributes of the instructors, we estimate the following regression:

(1) $(Very) \, Risky_{ic} = \beta X_i + \gamma W_c + u_{ic}$

---

[27] In the case of a tie, we used the following tie-breakers: 1. F2F is preferred over other modalities; 2. The class with a higher number of enrolled students is preferred; 3. The class in a smaller square footage room is preferred; 4. The lower-level course is preferred. We exclude online courses from this analysis as their safe capacity is undefined.

where $(Very)\ Risky_{ic}$ is an indicator that takes the value of one if class $c$ taught by instructor $i$ includes more registered students than (twice) the safe capacity prescribed by the CDC.[28]

The vector $X_i$ contains personal attributes of the instructor. For example, it includes dummies for the rank of the instructor within the University hierarchy. These dummies measure whether the instructor is a *GTA*, *Adjunct Instructor*, *Lecturer*, and *Staff*. The comparison category includes *Administrator*s, *Full Professor*s, *Associate Professors, or Assistant Professors*. Besides rank, we control for instructor's sex (*Female*), race (*Non-White*), *Age*, and tenure as an employee of the University (*Exp. Less Than 1 Year*, *Exp. 1-3 Years*, *Exp. 3-5 Years*, *Exp. 5-10 Years*). For the experience variable, the comparison group is the instructors with more than ten years of experience. In equation (1), $W_c$ stands for the class characteristics. These include the dummies for the level of the course (*100-*, *200-*, *300-*, *400- Level*, and *Master's Level*) and the modality of the class (*F2F*, *Mixed*, and *Unspecified*). The omitted categories are *Doctoral Level* and *Online* courses. We also control course type (lecture, lab, and so on) and the department/subject fixed effects (Economics, English, and so on). We cluster the standard errors at the instructor level.[29]

The results obtained from equation (1) are displayed in Table 5. In columns 1 and 2, the outcomes are *Risky* and *Very Risky*, respectively. In column 1, the coefficients of *GTA* and *Adjunct Instructor* are 0.066 and 0.085, respectively. This finding indicates that GTAs and adjunct instructors, who are ranked low within the University hierarchy, are about seven to nine percentage points more likely to teach a risky class compared to the tenure-track faculty (full, associate, and assistant professors) and administrators (such as the department chairs, deans, and others) who teach courses in the same department. We also tested whether these low-ranked instructors are more likely to teach risky classes than instructors of other ranks and found supporting evidence.[30] In other words, the adjunct instructors and GTAs are more likely to deliver a risky course compared to all instructors in their departments. Column 2 of Table 5, where the outcome is *Very Risky* (indicator for whether the number of registered students in a class is at least twice the CDC-prescribed safe capacity), provides similar results. GTAs and adjunct instructors are more likely to be recruited to teach very risky classes.

---

[28] We include only the classes whose safe capacity could be determined and online classes in the regressions.
[29] Result are robust to alternative clustering strategies, such as at the department or course level.
[30] The P-values for the test of equality of GTAs and Lecturers, and GTAs and Staff, are 0.10 and 0.04, respectively. The P-values for the test of equality of Instructors and other groups are both less than 0.02.

Among the other instructor attributes, *Female* stands out. Its positive and significant coefficients in Table 5 suggest that female instructors are more likely to teach risky classes. The negative coefficient of *Age* indicates that younger faculty are assigned to higher risk in their classrooms. More experienced instructors appear to be more likely to teach risky classes, but this relationship is not precisely estimated. Non-white and white instructors equally deliver undesirable courses. At the bottom of Table 5, we present the coefficients of class characteristics. The undergraduate courses, especially those in the 200-level, are riskier than the graduate-level classes. Not surprisingly, those with a *F2F* or *Mixed* modality are riskier than online courses.

*A Falsification Test with Classes Offered in Spring 2020*

The previous section shows that low-ranked instructors were teaching riskier classes than their tenured counterparts in the same departments in Spring 2021. This result could be due to the practice of the department chairs, who may have convinced lower-ranked instructors to deliver the risky courses that were unwanted by the tenured faculty. An alternative explanation could be the differences between the classes taught by the tenure-track versus the low-ranked instructors. For example, the tenure-track faculty typically have more education and experience than adjunct instructors and GTAs. As a result, the classes commonly taught by the tenure-track faculty may be inherently small and safe (perhaps specialized, higher-level, and field courses). In contrast, the classes assigned to the GTAs and the adjunct instructors may be large and essentially more risky classes (usually lower-level or introductory courses).

To test if classes taught by the lower-ranked instructors are inherently risky, we estimate equation (1) using Spring 2020 data when *none* of the classes were risky. This is because the COVID-19 pandemic was not widespread in the U.S. until mid-March, and when the pandemic became rampant, the university shut down the campus and switched to remote instruction. To operationalize our test, we computed the *hypothetical risk* that would have been observed if all classes were held in person.

Results obtained from equation (1) with the hypothetical risk in the Spring 2020 semester as the outcome variable are presented in Table 6. Although all of the control variables as in Table 5 are included in these regressions, in the interest of brevity, we provide the coefficients of only the variables that pertain to the personal attributes of instructors. The complete set of estimates is displayed in Appendix B Table B1. In striking contrast to the results shown in Table 5, the coefficients of variables *GTA* and *Adjunct Instructor* in columns 1 and 2 are close to zero and statistically insignificant. This finding indicates that GTAs, adjunct instructors, and tenure-track personnel (the comparison category) were equally likely to teach hypothetically risky or very risky classes in Spring 2020. We also tested if the coefficients of *GTA* and *Instructor* reported in column 1 are higher than other rank variables and found that they are not.[31] Table 6 shows that in Spring 2020, the low-ranked instructors did not deliver riskier courses than higher-ranked instructors, suggesting that the classes taught by these low-ranked instructors are not intrinsically risky. Estimates (presented in Appendix B Table B1) also indicate no relationship between the sex and age of an instructor and her/his tendency to teach a hypothetically risky class. Findings in Tables 5 and 6 suggest that instructors with specific attributes were systematically recruited in the Spring 2021 semester to teach riskier classes.[32]

**4.2. Is the Risk Compensated?**

In this section, using the monthly payroll records of the University employees, we test the hypothesis of whether instructors who taught riskier classes were also paid higher wages. We estimate the regression displayed below:

(2) $\quad Log(Wage_i) = \beta Risky_i + \alpha X_i + u_i$

The unit of observation in equation (2) is an instructor. $Log(Wage_i)$ represents the natural logarithm of the monthly earnings of instructor $i$ in Spring 2021. The variable of interest in equation (2) is $Risky$. This indicator variable equals one if the instructor's riskiest class in the semester includes more students than the safe capacity of the classroom. In other regressions, we use the variable *Very Risky*. This indicator takes the value of one if the number of registered students is at least twice as much as the safe capacity. Note that we exclude online courses from this analysis because their safe capacity is undefined.

Vector $X$ in equation (2) represents the individual and class level control variables. Most of these are the same as those in equation (1). For example, we control the instructor's rank within the University hierarchy, age, and experience as a university employee in the regressions. To capture the pay differences between instructors at different stages of their careers, we control for the complete set of interactions of rank, age, and experience variables in the regressions. Also included in *X* are the characteristics of the riskiest class taught by the instructor (dummies of the level, modality, lecture-type, and department fixed effects). Apart from these, we include the enrollment in the instructor's riskiest class, the area of the room where it is delivered, the numbers of undergraduate and graduate courses taught by the instructor in the semester, and the natural logarithm of the monthly wage of the instructor past year.[33] These additional variables are included in the regression because they may be correlated with an instructor's compensation and, simultaneously, with the variable of interest, *Risky*. For example, an instructor's pay may be higher if she/he teaches more classes, and the probability of teaching a risky class goes up with the number of courses. Similarly, an instructor could be paid a higher wage for delivering a large section of a class (with an increased number of students) or in a large space (e.g., in an auditorium).

---

[31] The P-values ranged between 0.60 and 0.81.
[32] We additionally checked if the teaching assignments were majorly altered. One thousand five hundred ninety-nine instructors taught at least one class in both the Spring 2020 and 2021 semesters. Just over 1,000 of them were tenure track faculty and administrators. For this group, about 60% saw changes in their classes in the Spring 2021 semester *and* taught in safe classrooms. This share is only 37% for the lower-ranked personnel, indicating a reshuffling of class assignments and COVID-19 risk to the disadvantage of lower-ranked instructors.

*Instrumental Variables Estimation*

The primary threat to identification is the potential endogeneity of the *Risk* variable in equation (2) which may bias our estimates upward or downward.[34] For example, omitting from our regressions the extent of the collegiality of an instructor may lead to an upwards bias. This is because, to avoid conflict with her/his department chair or the university administration, a collegial instructor would act more sympathetic toward the university's push for more in-person classes. She/he would have a greater tendency or volunteer to teach risky classes. At the same time, collegial instructors could be rewarded with higher wages by the university independent of their propensity to teach risky courses relative to their "troublemaker" counterparts.[35] Alternatively, endogeneity may cause a downward bias due to information asymmetries. For example, suppose an instructor is unaware of or unable to predict the actual risk of teaching. In that case, the department chair may assign her/him a risky class without paying a wage premium.[36] We implement an IV strategy to guard against these threats.

---

[33] We predicted the square footage of the few classrooms for which it was not listed on the CCAs based on the number of seats in the room and the building in which it was located.
[34] For completeness, we present the OLS estimates obtained from equation (2) in Appendix B Table B2.
[35] Collegiality is a condition for tenure and promotion under the policies of Auburn University.

*The Instrument: Can the Students Be Dispersed While Attending the Class?*

Our instrument, which measures whether students can be dispersed when attending a lecture, is constructed based on whether the classrooms' furniture is fixed to the ground or easily movable. Examples of fixed and movable furniture are presented in Appendix A Figures A5 and A6. The classroom photographed in Appendix A Figure A5 is Room 230 in Miller Hall, where the Department of Economics is located at Auburn University. The seats in this classroom are planted to the ground, and the desks are attached to the seats. As a result, a student cannot move her/his desk from its location within this room, making it an example of a fixed-furniture classroom. Appendix A Figure A6 exhibits a view of Room 207 in the same building. The seats and desks in this classroom have wheels underneath them, enabling students to move them. Room 207 is an example of a movable furniture classroom.

Whether students can be dispersed in the room thanks to the furniture in it and layout has a mechanical relationship with the endogenous variable, *Risky*, which is computed as a function of the ratio of the number of registered students to the classroom's safe capacity (the number of students who can attend the class and still maintain CDC-prescribed six-foot social distancing). Other things, such as the square footage of the classroom, same, *classrooms with movable furniture have a larger safe capacity compared to rooms with fixed furniture.*[37] This is because, with moveable furniture, students can more easily transport themselves to a farther location in the classroom to keep a six-foot social distance. For example, rooms 207 and 230 in Miller Hall are similar in square footage (approximately 1,300 and 1,200 square feet, respectively). The safe capacity in room 207, which has movable furniture, is 23, while in Room 230, which is a fixed classroom, only 15 students can fit safely, per the calculations in the CCAs of the University (see Appendix A Figures A2 and A3).

---

[36] It is likely that some instructors were misinformed about the risk of their classes, as the safe capacities of the classrooms were not made available to the campus community until December 18, 2020, long after the teaching schedules were finalized in November. In addition, these studies were not advertised until January 2021.
[37] The director of facilities at Auburn University confirmed this relationship in a public meeting of the university. Appendix A provides more details.

We also have statistical evidence that corroborates the positive relationship between a classroom's safe capacity and furniture type. We estimate a regression where the unit of observation is a classroom (N=381). To compare rooms of equal size, we also control for the square footage of the space in this regression. The outcome is the CDC-prescribed safe capacity of the room. On the right-hand side, we have the variable *Dispersible Class*, which is an indicator for whether all students in a classroom can spread away from each other while attending the lectures, or put differently, if that classroom has movable furniture, and zero otherwise. The coefficient of *Dispersible Class* is about 4.5 and is statistically significant, indicating that in classrooms with movable furniture, in which students can move their seats and desks, four to five additional students can fit in safely compared to fixed-furniture classrooms.

*Validity of the IV Strategy*

The internal validity of our IV strategy depends on whether four assumptions hold. The first of these assumptions is <u>Relevance</u>, which requires the instrument to be a strong predictor of the endogenous variable, conditional on other control variables. We tested our instrument's relevance by estimating the first-stage regression, which takes the following form:

(3) $\quad Risky_i = \beta Dispersible\ Class_i + \alpha X_i + u_i$

Our instrument, $Dispersible\ Class_i$, takes the value of one if students registered in instructor $i$'s riskiest class can spread away from each other when attending the lecture. The outcome variable in equation (3) is *Risky*, and in some regressions, we use *Very Risky* instead. Vector $X_i$ represents the controls. It includes all control variables used in equation (2).

Table 7 displays the estimates of $Dispersible\ Class$ in equation (3). The coefficients of the complete set of control variables are presented in Appendix B Table B3. The coefficient of *Dispersible Class* is -0.198 in column 1, indicating that such classes, where the classroom setting allows students to outspread from one another, are about twenty percentage points less likely to be risky. Notably, the F-statistic for the significance of the instrument is approximately 35, suggesting a strong relationship between our IV and the endogenous variable.[38] Thus, our IV strategy is unlikely to suffer the consequences of weak I.V.s (e.g., see Bound, Jaeger, and Baker 1995). Column 2 presents a similar result for *Very Risky*. Dispersible classes are about fifteen percentage points less likely to be very risky (F-statistic: 29). The evidence in Table 7 suggests that our instrument is strong and shifts the endogenous variable in the expected direction.

---

[38] Recent research has shown the first-stage F-statistics should be over a 100 to generate zero-distortion confidence intervals (Lee et al. 2020). In auxiliary analyses shown in Appendix B Table B7, where we consider a continuous risk measure, *Classroom Risk*, the first-stage F-statistic is close to this threshold (90.97). See the notes to that table for further details.

The second assumption for the internal validity of an IV strategy is <u>Monotonicity</u>, which requires the endogenous variable to be moved only in one direction by the instrument. Although this assumption is not empirically testable alone, we studied its plausibility by analyzing the variables underlying its construction. Recall that the endogenous variable *Risky* is generated using the ratio of the number of registered students to the safe capacity of the room, in short, *Classroom Risk*. Figure 1 depicts this ratio's cumulative distributions (cdf) for the samples of classes with *Dispersible Class* equal to one and zero, separately. The thin red (thick blue) line represents (non-) dispersible classes. As evident in Figure 1, these cdfs do not intersect, implying that the monotonicity assumption is likely to hold, per Angrist and Imbens (1995).[39]

The third assumption is instrument <u>Excludability</u>, which requires no relationship between the instrumental variable and the outcome except through the endogenous variable, conditional on other control variables. In our context, the exclusion restriction requires that whether an instructor teaches a *Dispersible Class* should not affect her/his wages indirectly other than through the riskiness of her/his classes. This assumption is unlikely to be violated in our setting. Nonetheless, we include many control variables in the hedonic wage regressions. For example, we control for not only a flexible function of the personal attributes of the instructor but also class characteristics that may influence her/his wages, such as the level of the class, modality, the number of students, and the square footage of the room. To avert the possibility that specific departments or lecture types require a particular kind of classroom, we control department and lecture type fixed effects in the regressions.

The last assumption necessary for the internal validity of our IV strategy is <u>Independence</u>. This condition requires the instrument to be independent of potential outcomes (wages) and potential treatment (riskiness of the class). This assumption is also not testable since the counterfactuals cannot be observed. Instead, we study the correlation between the instrument and the observable class and instructor characteristics by regressing each instructor's individual and class attributes on *Dispersible Class*. To compare rooms of equal size, we control the area of the room. We also control for the department fixed effects in these regressions. The estimates are given in Figure 2, in which each row represents a separate regression. The dots mark the standardized point estimates, and the lines are the 95% confidence intervals.[40] Results indicate that the attributes of the instructors and their riskiest classes are not statistically different between dispersible versus non-dispersible classes.[41] This finding reinforces the idea that whether a class is dispersible is as good as random. Thus, our instrument is independent of potential outcomes.

---

[39] Suppose we consider the continuous measure of risk, *Classroom Risk* (denoted by $r$) and define the potential risk of a class that is (not) dispersible by $Risk_0$ ($Risk_1$). Then the monotonicity assumption requires $Prob(Risk_1 - Risk_0 \geq 0) = 1$ or $P(Risk_1 \geq r) \geq P(Risk_0 \geq r)$. If our instrument ($D$) is independent of potential risk as well (for which we provide evidence below), then the monotonicity assumption implies $P(Risk_1 \geq r \mid D = 1) \geq P(Risk_0 \geq r \mid D = 0)$ or $F(r \mid D = 1) \geq F(r \mid D = 0)$, where $F()$ is the cdf of risk.

*Instrumental Variables Estimates*

To operationalize our IV strategy, we implement the Two-Stage Least Squares method. The estimates for the variables of interest obtained from this exercise are presented in Table 8.[42] The coefficient of *Risky* in column 1 is 0.175, and it is statistically different from zero at the 5% level. This result indicates that instructors who teach at least one risky class earn approximately 20% more than their counterparts who deliver only safe course sections. Similarly, in column 2, the estimate of *Very Risky* is 0.228, indicating a wage premium of 26%. Relative to the average monthly wage of an instructor in our sample, these effects correspond to approximately $1,850 per month or $7,400 in a four-month semester.[43] This is our estimate of the price of teaching at least one risky class. It corresponds to about $5,000 *per class* over the four-month semester.[44] In column 3, we present the reduced form estimates. Instrument's coefficient in this regression is -0.035.[45]

*A Placebo Test With Spring 2020 Payroll Records*

---

[40] We present the raw coefficients and P-values, adjusted for multiple hypothesis testing, in Appendix B Table B4.
[41] We also estimated the same regression with the average of the classroom characteristics as the outcome at the classroom level. The coefficient of *Dispersible Class* was always insignificant at usual confidence levels.
[42] The full set of estimates are presented in Appendix B Table B5.
[43] We computed this figure by using the mean monthly salary of instructors in Table 4 ($9,669). The 19.12% premium implies about $1,850 extra earnings each month.
[44] The expected number of risky classes taught, conditional on teaching at least one risky class, is 1.5 We computed this by examining the sample of instructors who taught more than one class in the Spring 2021 semester. For 41% of these instructors, their riskiest class is *Risky*. For half of these instructors, their second riskiest class was also *Risky*. Thus, the per-class wage premium is $7,400 divided by 1.5, or about $5,000.
[45] A possible concern is that very large classes with hundreds of students are driving our baseline results. To check if this is the case, we re-estimated the 2SLS regressions using the sample of riskiest classes where the number of students is less than 100 (N=1,419). The results are similar to those shown in Table 8.

One may argue that risky classes also have attributes other than a higher risk of COVID-19, such as being more crowded and congested. Under this premise, the wage premium we estimated may not be for the classroom's COVID-19 risk but for these additional attributes. In this section, we test whether this premise is correct. For this analysis, we used data from Spring 2020, when none of the classes were risky. Under the hypothesis that instructors are paid for the COVID-19 risk but not for other class attributes, there should be no correlation between the hypothetical risk and pay in Spring 2020.

We estimated a regression that is identical to equation (2), except the variable *Risky* is replaced with *Hypothetically Risky*. This variable takes the value of one if the class taught by the instructor in Spring 2020 would have been classified as risky if it was delivered in Spring 2021. The 2SLS results obtained from this specification are presented in Table 9.[46] Although the reduced-form relationship between our instrument and the endogenous variable is strong, *Hypothetically Risky Class*'s coefficients are small, negative, and statistically insignificant in both regressions.[47] The evidence in this section suggests that instructors were not paid a premium for teaching congested or more crowded courses.

*Alternative Measurements of Risk*

Although we possess a continuous risk measure (the ratio of the number of students in a class to the safe and lawful capacity of the room), we chose to use an indicator treatment in our empirical application. This is because our instrument takes a binary form, and the "0-to-1" treatment margin for risk is more intuitive. On the downside, our treatment variable can only differentiate between the risky from the safe classes, i.e., whether a course's risk is above a threshold. One may be concerned about our treatment's inability to take into account the risk differences between two Risky (or Safe) classes and consequent misspecification problems.

We address this issue in two ways. First, we re-estimate our wage regression (equation 2) using risk variables constructed based on different thresholds. Specifically, the variable $Risky^t$ equals to one if the ratio of the number of students registered for the class to the room's safe capacity is greater than threshold $t$, and zero otherwise.[48] $Risky^1$ and $Risky^2$ variables are the same as the $Risky$ and $Very\ Risky$ variables we used in our primary analysis. We run a series of 2SLS regressions with $Risky^t$, considering $t = 0.5, 0.6, 0.7, \ldots$. The results are presented in Figure 3, where the coefficients of $Risky^t$ (solid blue line) and 95% confidence bands (light blue dashed lines) are displayed. The vertical lines mark thresholds for one and two. Figure 3 shows that the impact of being in a risky class on log wages is stable at around 0.2 and increases slightly for higher risk thresholds. The standard errors inflate for extremely high levels of risk, possibly because there are few such classes (about 8% of the instructors' riskiest classes have a risk ratio greater than 3). Alternatively, this could be caused by the weaker relationship between our instrument and $Risky^t$ for large t. See Appendix B Figure B1 for the first stage results.

---

[46] The complete set of estimates are in Appendix B Table B6.
[47] The coefficients of *Dispersible Class* in the first stage regressions were -0.127 and -0.123, for *Hypothetically Risky* and *Hypothetically Very Risky*, respectively. The first-stage F-statistics are 22.26 and 16.83.

We also estimate our wage regressions using the continuous measure of risk. We estimate a specification identical to equation 2, except we replace the indicator treatment variable *Risky* with the continuous variable *Classroom Risk* (mean: 1.12). The results are presented in Appendix B Table B7. The 2SLS estimate of the effect of continuous *Classroom Risk* on log wages (column 3) is 0.076, and it is statistically significant at the 5% level. This finding implies that a one percent increase in *Classroom Risk* causes wages to increase by 0.09 percent.[49]

## 5. Did Auburn Save by Ignoring CDC's Social Distancing Guidelines?

We will now discuss the financial motivations behind Auburn University's conduct in Spring 2021. This semester, possibly due to the pressures from politicians, the local businesses, and the parents and students, Auburn University tried to provide as many in-person classes as possible without complying with the CDC-prescribed 6 foot social distancing in the classrooms, which was also the Alabama Governor's order. As a result of these university practices, at least 1,379 classes were delivered in risky classrooms by 771 instructors. Our estimates indicate that the cost to the university of teaching at least one risky class is approximately $7,400, bringing the total bill to about $5.7 million (771 × $7,400).

Next, we calculate the hypothetical cost that Auburn University would have incurred if it conformed to the CDC guidelines and the Governor's public health order. Notably, in this scenario, the university still provides as much in-person instruction as it did, but all in-person classes are delivered in safe rooms. Our analysis shows that the university needed to open at least 2,246 additional sections for the complete set of classes in Spring 2021 to be safe.[50] The number of required additional sections varies by department. To account for the differences in the market wages of instructors in different departments, we use the average pay per class using the payroll data of adjunct instructors within each department.[51] We find that Auburn University would have had to spend $13.5 million on the instructors in total.

---

[48] Say, the number of registered students for a class is 35 and the safe capacity of that class is 30. The ratio is 35/30=1.16. Thus, for this class, $Risky^{1.1}$ variable is equal to one, but $Risky^{1.2}$ is zero.

[49] We computed this elasticity by multiplying the coefficient of Classroom Risk in column 3 of Appendix B Table B7, 0.076, and the sample mean of the *Classroom Risk*, 1.12.

The $7.8 million difference ($5.7 – $13.5) between what Auburn actually incurred to provide as much in-person instruction as it did in Spring 2021 versus what Auburn could have spent and provided the same amount of in-person instruction, but in a safer way, constitutes a lower bound for the university's cost savings. This is because the $13.5 million figure includes only the labor costs, ignoring other potential costs (the additional overhead or utility costs or compensating differentials associated with teaching overloads and class times outside business hours).

## 6. Summary and Discussion

Universities are often an essential source of economic growth, especially in small college towns where the financial health of the local businesses and universities depends on the physical presence of students (see, e.g., Lane and Johnstone 2012; Goldstein and Drucker 2006). During the early days of the COVID-19 pandemic, when the universities switched to remote operations and the students left, the local economies of college towns and the university budgets suffered. The potentially adverse effects of remote instruction on students' learning and career trajectories (e.g., see Jaeger et al. 2021; Aucejo et al. 2020) imposed further pressures on the universities to reopen. As a result, university administrators and other stakeholders pushed for more in-person operations in the following academic year, which required reducing or eliminating safety risks associated with viral transmission in the university community.

---

[50] To obtain this number, for each class, we calculated the ratio of registered students to the safe capacity of that class, rounded up to the next integer, and subtracted one. For example, if in a class 58 students are registered, and the safe capacity of the classroom is 35, then the ratio is 58/35=1.66. This indicates at least one more section of the same class is required to conform to six-foot social distancing guideline.

[51] For example, an Economics adjunct instructor is paid about $7,800 per class, while an instructor in History is paid $5,000, on average.

The experience of Auburn University (a large public university in Alabama) was no different.[52] Possibly facing political and financial pressures and high demand for in-person instruction, the university reopened in the 2020-2021 academic year, yet without taking all precautions required by law for the resumption of face-to-face classes. Specifically, despite the CDC's guidance and a public health order from the Governor of Alabama to the contrary, the Auburn University administration did not implement a social distancing policy in the classrooms. As a result, at least half of the face-to-face classes were taught in classrooms with enrollments that prohibit a six-foot distancing when all registered students attended.

Our results show that graduate student teaching assistants (GTAs) and non-tenure-track adjunct instructors were systematically recruited over instructors who hold higher ranks (i.e., tenure track professors) to teach the risky classes in Spring 2021. We also estimate whether the instructors who delivered risky classes are compensated using an instrumental variables strategy, where the risk of a course is shifted by the type and setup of the furniture in classrooms. The classes that have movable desks and chairs or those that otherwise allow students to disperse away from one another during lectures are less risky. We provide empirical evidence for the validity and strength of this instrument. Our "price" estimate of teaching at least one risky class (over teaching only safe classes) is approximately $7,400 in a four-month semester.

Because we estimate a risk premium for COVID-19, our paper can be compared to the literature on compensating wage differentials.[53] However, the market for instructional services during the reopening of Auburn University does not comport with a traditional setting of workers sorting based upon risk preferences and bargaining with employers over the risk premia. Instead, the risk premium of $7,400 is best viewed as the average valuation of risk by Auburn's central administration, who minimizes the costs of delivering in-person classes. We also contribute to the literature in the study of institutions, such as Acemoglu and Robinson (2008), as we study the

---

[52] In fact, in many ways, Auburn's experience during the pandemic mimics the larger political and economic dislocations of the rest of the United States. The lack of coordinated action, together with the university's aggressive push for more in-person instruction in Spring 2021, led to a call for a "vote of no confidence" in the Provost at a general faculty meeting, in the Fall semester of 2020. See Appendix A for more details.

[53] The most common is the study of the impact on wages of occupational hazards (Lavetti 2020; Lavetti and Schmutte 2018; Kniesner, Viscusi, Woock and Ziliak 2012; Cousineau, Lacroix and Girard 1992; Viscusi and Moore 1991; Garen 1988). In a very recent paper, Wissmann (2022) shows smoking bans in restaurants and bars in Germany caused a decrease in earnings of the waitresses/waiters. Others focus on a range of non-market goods, such as sexual harassment (Hersch 2011), work shifts (Kostiuk 1990), schedules (Mas and Pallais 2017), commute time (Mulalic, Van Ommeren and Pilegaard 2014), health insurance benefits (Qin and Chernew 2014), and contract guarantees (Link and Yosifov 2012).

disaggregated behavior of elites in the allocation and compensation of health risk. Consistent with past research, our findings show that elites' preferences are expressed in the aggregate outcome in extreme risk situations, such as those involving life and death (Frey, Savage, and Torgler 2011).[54]

Our back-of-the-envelope calculations show Auburn University saved about $8 million in labor costs in Spring 2021 by ignoring the CDC-prescribed social distancing guidelines. Was Auburn University's conduct in Spring 2021 socially optimal? The answer depends on whether external costs to society exceed these cost savings. It is plausible that the university's actions affected the students who were not compensated for the COVID-19 risk they had borne. Due to attending risky classes, they may have been exposed to the virus and suffered illness. In addition, students in the risky classes may have contributed to the further spread and mutation of the COVID-19 virus. These considerations suggest that Auburn University's conduct in Spring 2021 may not be socially optimal despite being justifiable based on its cost savings.[55]

---

[54] Other recent studies demonstrate that the distribution of occupational risk associated with COVID-19 is skewed toward females and other disadvantaged groups (Baylis et al., 2020; Mongey, Pilossoph, and Weinberg 2020; Hawkins 2020a; 2020b; Tai et al. 2021; Yancy 2020).

[55] An additional discussion is around whether Auburn could have provided a safe in-person learning environment without spending more than the budget allowed. For this, the university had to cut other expenditures such as the raises given to the employees in 2021 (about $13 million).

**References**

Acemoglu, Daron, and James A Robinson. (2008). "Persistence of Power, Elites, and Institutions." *American Economic Review* 98(1), pp. 267–93.

Almond, Douglas. (2006). "Is the 1918 Influenza Pandemic over? Long-Term Effects of in Utero Influenza Exposure in the Post-1940 US Population." *Journal of Political Economy* 114(4), pp. 672-712.

Altindag, Duha Tore, Elif S. Filiz, and Erdal Tekin. (2021). "Is Online Education Working?" NBER Working Paper Series #29113.

Aucejo, Esteban M., Jacob French, Maria Paola Ugalde Araya, and Basit Zafar. (2020). "The Impact of COVID-19 on Student Experiences and Expectations: Evidence from a Survey." *Journal of Public Economics* 191(November): 104271.

Baylis, Patrick, Pierre-Loup Beauregard, Marie Connolly, Nicole Fortin, David A. Green, Pablo Gutierrez Cubillos, Sam Gyetvay, Catherine Haeck, Timea Laura Molnar, and Gaëlle Simard-Duplain. (2020). "The Distribution of COVID-19 Related Risks." NBER Working Paper Series #27881.

Bird, Kelli A., Benjamin L. Castleman, and Gabrielle Lohner. (2020). "Negative Impacts From the Shift to Online Learning During the COVID-19 Crisis: Evidence from a Statewide Community College System." EdWorkingPaper No. 20-299. Annenberg Institute for School Reform at Brown University.

Bound, John, David A Jaeger, and Regina M Baker. (1995). "Problems with Instrumental Variables Estimation When the Correlation between the Instruments and the Endogenous Explanatory Variable Is Weak." *Journal of the American Statistical Association* 90(430), pp. 443-50.

Cousineau, Jean-Michel, Robert Lacroix, and Anne-Marie Girard. (1992). "Occupational Hazard and Wage Compensating Differentials." *The Review of Economics and Statistics* 74(1), pp. 166-169.

Ehrlich, I., & Becker, G. S. (1972). Market Insurance, Self-Insurance, and Self-Protection. Journal of Political Economy, 80(4), 623-648.

Frey, Bruno S, David A Savage, and Benno Torgler. (2011). "Behavior under Extreme Conditions: The Titanic Disaster." *Journal of Economic Perspectives* 25(1), pp. 209-22.

Garen, John. (1988). "Compensating Wage Differentials and the Endogeneity of Job Riskiness." *The Review of Economics and Statistics* 70(1), pp. 9-16.

Goldstein, Harvey, and Joshua Drucker. (2006). "The Economic Development Impacts of Universities on Regions: Do Size and Distance Matter?" *Economic Development Quarterly* 20(1), pp. 22-43.

Hawkins, Devan. (2020a). "Differential Occupational Risk for COVID-19 and Other Infection Exposure According to Race and Ethnicity." *American Journal of Industrial Medicine* 63(9), pp. 817-20.

Hawkins, Devan. (2020b). "Social Determinants of COVID-19 in Massachusetts, United States: An Ecological Study." *Journal of Preventive Medicine and Public Health* 53(4), pp. 220-227.

Hersch, Joni. (2011). "Compensating Differentials for Sexual Harassment." *American Economic Review* 101(3), pp. 630-34.

Jaeger, David A., Jaime Arellano-Bover, Krzysztof Karbownik, Marta Martínez-Matute, John M. Nunley, Alan Seals, Mackenzie Alston, et al. (2021). "The Global COVID-19 Student Survey: First Wave Results." SSRN Scholarly Paper ID 3860600.

Kniesner, Thomas J., W. Kip Viscusi, Christopher Woock, and James P. Ziliak. (2012). "THE VALUE OF A STATISTICAL LIFE: EVIDENCE FROM PANEL DATA." *The Review of Economics and Statistics* 94(1), pp. 74-87.

Kofoed, Michael S., Lucas Gebhart, Dallas Gilmore, and Ryan Moschitto. (2021). "Zooming to Class?: Experimental Evidence on College Students' Online Learning during COVID-19." IZA Discussion Papers 14356.

Kostiuk, Peter F. (1990). "Compensating Differentials for Shift Work." *Journal of Political Economy* 98(5), pp. 1054-1075.

Lane, Jason E., and D. Bruce Johnstone, eds. (2012). *Universities and Colleges as Economic Drivers: Measuring Higher Education's Role in Economic Development.* SUNY Series, Critical Issues in Higher Education.

Lavetti, Kurt. (2020). "The Estimation of Compensating Wage Differentials: Lessons From the *Deadliest Catch*." *Journal of Business & Economic Statistics* 38(1), pp. 165-182

Lavetti, Kurt, and Ian Schmutte. (2018). "Estimating Compensating Wage Differentials with Endogenous Job Mobility." *Mimeo*. Accessed at <http://kurtlavetti.com/CDEM_vc.pdf>

Lee, David L., Justin McCrary, Marcelo J. Moreira, and Jack Porter. 2020. "Valid T-Ratio Inference for IV." ArXiv Preprint ArXiv:2010.05058.

Link, Charles R., and Martin Yosifov. (2012). "Contract Length and Salaries Compensating Wage Differentials in Major League Baseball." *Journal of Sports Economics* 13(1), pp. 3-19.

Loertscher, S., & Marx, L. M. (2022). "Incomplete information bargaining with applications to mergers, investment, and vertical integration." *American Economic Review* 112(2), 616-49.

Mas, Alexandre, and Amanda Pallais. (2017). "Valuing Alternative Work Arrangements." *American Economic Review* 107(12), pp. 3722-59.

Mongey, Simon, Laura Pilossoph, and Alex Weinberg. (2020). "Which Workers Bear the Burden of Social Distancing Policies?" NBER Working Paper Series #27085.

Mulalic, Ismir, Jos N. Van Ommeren, and Ninette Pilegaard. (2014). "Wages and Commuting: Quasi-natural Experiments' Evidence from Firms that Relocate." *The Economic Journal* 124(579), pp. 1086-1105.

Qin, Paige, and Michael Chernew. (2014). "Compensating wage differentials and the impact of health insurance in the public sector on wages and hours." *Journal of Health Economics* 38(December), pp. 77-87.

Rosen, S., 1986. The theory of equalizing differences. *Handbook of Labor Economics*, 1, pp.641-692.

Tai, Don Bambino Geno, Aditya Shah, Chyke A. Doubeni, Irene G. Sia, and Mark L. Wieland. (2021). "The Disproportionate Impact of COVID-19 on Racial and Ethnic Minorities in the United States." *Clinical Infectious Diseases* 72(4), pp. 703-706.

Viscusi, W. Kip, and Michael J. Moore. (1991). "Worker Learning and Compensating Differentials." *Industrial and Labor Relations Review* 45(1), pp. 80-96.

Wissmann, Daniel. 2022. "Finally a Smoking Gun? Compensating Differentials and the Introduction of Smoking Bans." American Economic Journal: Applied Economics, 14 (1): 75-106.

Yancy, Clyde W. (2020). "COVID-19 and African Americans." *JAMA* 323(19), pp. 1891-1892.

**Figure 1**
**Cumulative Distribution of *Risk* by the Status of the Instrument**

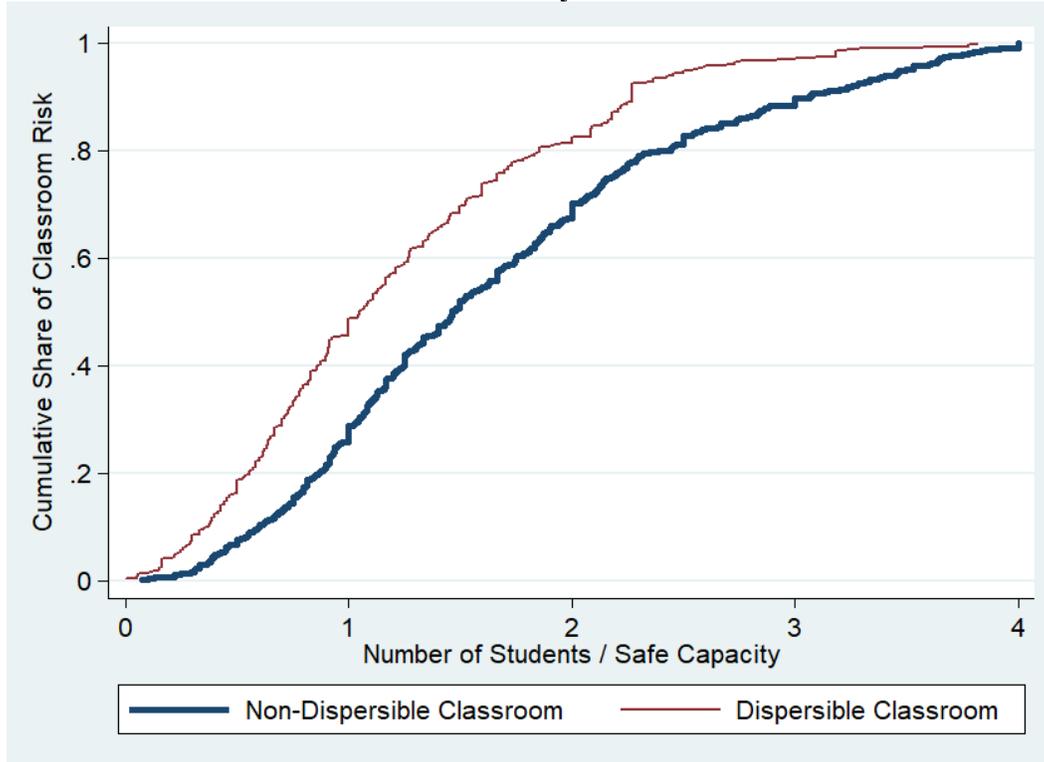

The figure displays the cumulative share of the ratio of the number of registered students to the safe capacity of the room, in short, *risk*, for the samples of each instructor's riskiest class with *Dispersible Class* is equal to one and zero (N=1,501).

**Figure 2**
**Correlation Between the Instrument and Observable Characteristics**

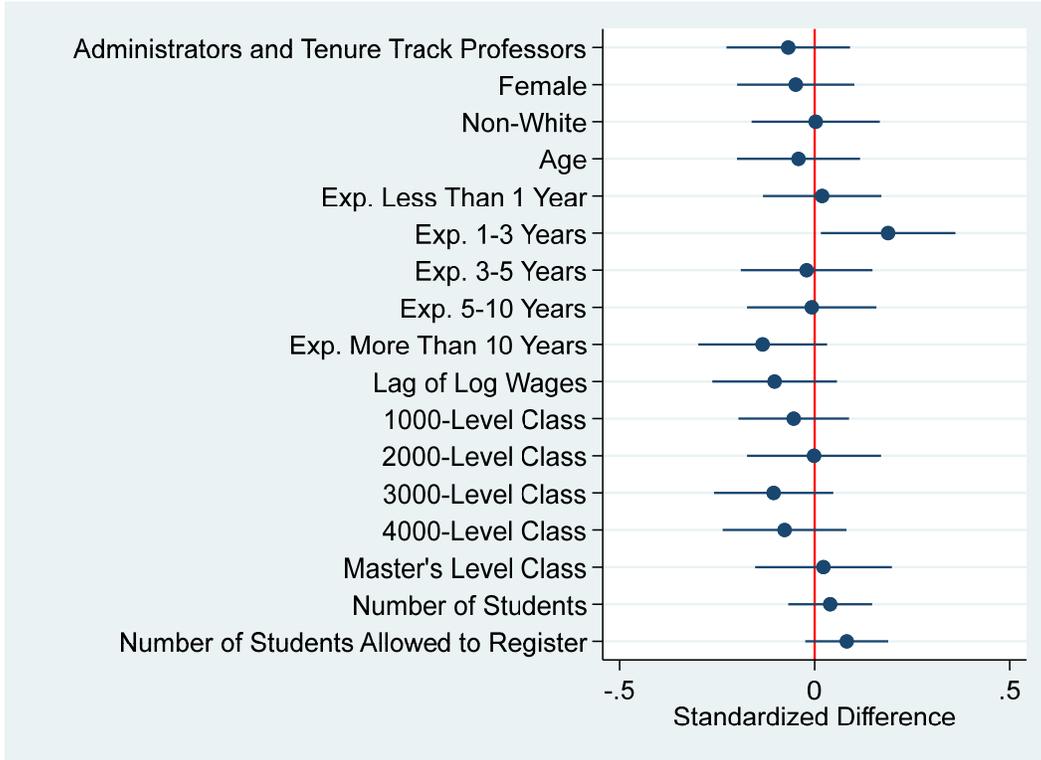

Each row represents the (standardized) estimates of *Dispersible Class*, where characteristics of the instructors and the number of students in their riskiest class characteristics are the outcomes (N=1,501). Regressions additionally control for the department fixed effects and classroom area. The dots mark the standardized point estimates, and the lines are the 95% confidence intervals. The vertical red line indicates zero.

**Figure 3**
**Marginal Treatment Effects**

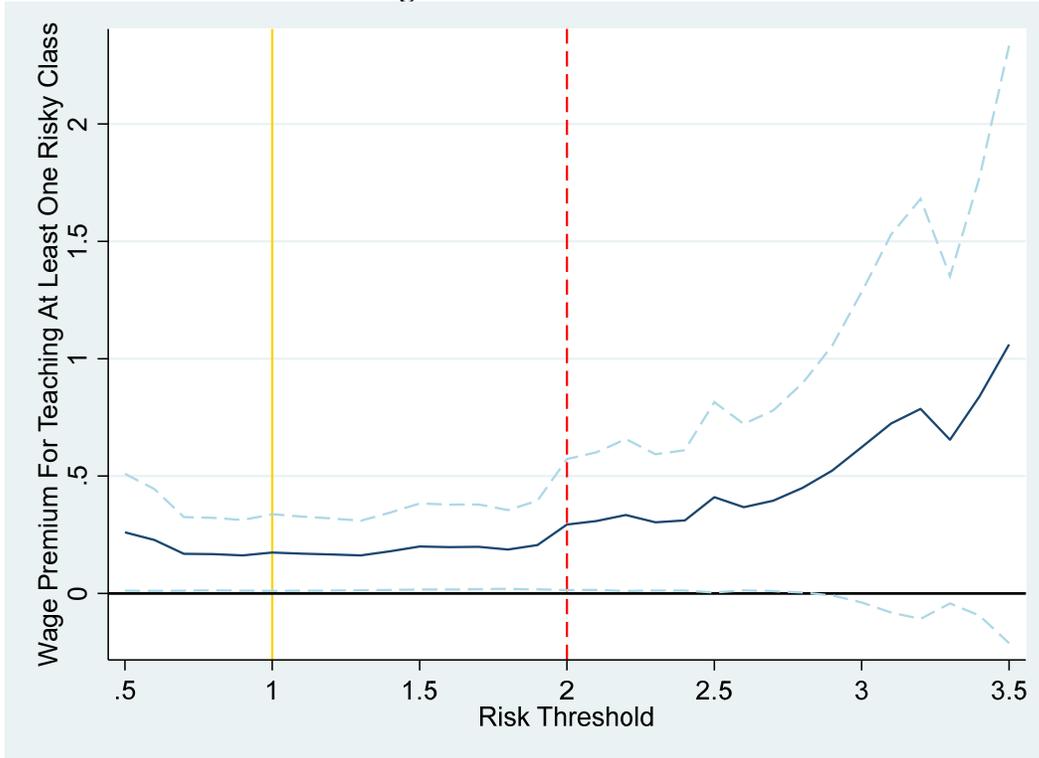

The solid blue line represents 2SLS point estimates for the relationship between classroom risk and instructor pay. Dashed lines represent the 95% confidence intervals. Estimates are from eq. (2), where $Risky^t$ is equal to one if the ratio of the number of students in the class to the CDC safe capacity of the room exceeds the risk threshold ($t$) denoted on the horizontal axis. All risk thresholds between 0.5 and 3.5 are considered, in increments of 0.1. The solid yellow and dashed red vertical lines indicate $t$ equals to one and two, respectively, corresponding to variables *Risky* and *Very Risky*.

**Table 1**
**Distribution of Classes by Modality in 2020-2021 Academic Year**

|              | Fall 2020 | Spring 2021 |
|--------------|-----------|-------------|
| N: # classes | N=7,469   | N=7,017     |
| F2F          | 19.17%    | 70.67%      |
| Mixed        | 47.68%    | 5.53%       |
| Online       | 33.15%    | 23.80%      |

F2F is the share of all Face-to-Face classes. Online courses are those which take place solely online. Mixed courses employ a mix of in-person and online formats.

**Table 2**
**Risk in Classes at Auburn University in Spring 2021**

|                                | F2F     | Mixed  | Unspecified | Online   |
|--------------------------------|---------|--------|-------------|----------|
| N: # classes                   | N=2,531 | N=195  | N=96        | N=1,560  |
| Safe Classes                   | 51.05%  | 40.0%  | 76.04%      | 100.00%  |
| Risky but not Very Risky Classes | 26.63%  | 31.79% | 12.50%      | 0.00%    |
| Very Risky Classes             | 22.32%  | 28.21% | 11.46%      | 0.00%    |

Only the classes that can be classified according to their risk are included in the analysis. Safe Classes have fewer students than the CDC-prescribed safe capacity, i.e., the maximum number of students who can be seated while maintaining a six-foot distance between. Risky Classes have more students than the safe capacity. Very Risky Classes include more students than twice the safe capacity.

**Table 3**
**Descriptions and Summary Statistics of Class Attribute Variables in Spring 2021**

| Variable | Description | Mean | Std. Dev. |
|---|---|---|---|
| No. Students | The number of students registered for the class | 26.11 | 36.71 |
| Safe Class Capacity | The maximum number of students that can be seated in the classroom with six-foot social distancing | 21.14 | 14.09 |
| Classroom Area | The area of the classroom in thousands of square feet | 1.46 | 0.93 |
| Safe Class | =1 if the ratio of No. Students to Safe Class Capacity is less than or equal to one. | 0.69 | 0.46 |
| Risky Class | =1 if the ratio of No. Students to Safe Class Capacity is greater than one. | 0.31 | 0.46 |
| Very Risky Class | =1 if the ratio of No. Students to Safe Class Capacity is greater than or equal to two. | 0.14 | 0.35 |
| F2F | =1 if the class is delivered in person. | 0.58 | 0.49 |
| Mixed | =1 if the class is delivered with a mix of in-person and online modalities. | 0.04 | 0.21 |
| Online | =1 if the class is delivered in a purely online modality | 0.36 | 0.48 |
| Unspecified Modality | =1 if the class modality is not specified. | 0.02 | 0.15 |
| 1000-Level | =1 if the class level is 1000. | 0.23 | 0.42 |
| 2000-Level | =1 if the class level is 2000. | 0.16 | 0.37 |
| 3000-Level | =1 if the class level is 3000. | 0.19 | 0.39 |
| 4000-Level | =1 if the class level is 4000. | 0.11 | 0.31 |
| Master's Level | =1 if the class is Master's level. | 0.16 | 0.36 |
| Doctoral Level | =1 if the class is Doctoral level. | 0.16 | 0.37 |
| Lecture Type | The primary means of instruction for the class | 18 categories | |
| Department | The four-letter subject code for the class | 143 categories | |

The unit of observation is a class. Only the classes that can be classified according to their risk are included in the analysis. The number of observations is 4,382.

**Table 4**
**Descriptions and Summary Statistics of Instructor Attribute Variables in Spring 2021**

| Variable | Description | Mean | Std. Dev. |
|---|---|---|---|
| *Individual Characteristics* | | | |
| Female | =1 if the instructor is female. | 0.40 | 0.49 |
| Non-White | =1 if the instructor's race is not White. | 0.22 | 0.41 |
| Administrator | =1 if the instructor also has an administrative position, such as department chair, dean, or provost. | 0.07 | 0.26 |
| Full Prof. | =1 if the instructor is a full professor. | 0.21 | 0.40 |
| Assoc. Prof. | =1 if the instructor is an associate professor. | 0.21 | 0.41 |
| Assist. Prof. | =1 if the instructor is an assistant professor. | 0.21 | 0.41 |
| Lecturer | =1 if the instructor is a lecturer. | 0.08 | 0.26 |
| Staff | =1 if the instructor also works in a staff position at the university. | 0.06 | 0.24 |
| Adjunct Instructor | =1 if the instructor is a contract instructor. | 0.07 | 0.26 |
| GTA | =1 if the instructor is a graduate student. | 0.09 | 0.28 |
| No. Undergrad. Classes | The number of undergraduate classes taught by the instructor | 2.06 | 3.90 |
| No. Grad. Classes | The number of graduate classes taught by the instructor | 1.83 | 2.39 |
| Monthly Wage | Monthly pay of the instructor in Spring 2021. | $9669.45 | 5369.21 |
| Exp. Less Than 1 year | =1 if the instructor has less than one year of experience. | 0.01 | 0.10 |
| Exp. 1-3 Years | =1 if the instructor has one to three years of experience. | 0.19 | 0.40 |
| Exp. 3-5 Years | =1 if the instructor has three to five years of experience. | 0.15 | 0.36 |
| Exp. 5-10 Years | =1 if the instructor has five to ten years of experience. | 0.22 | 0.41 |
| Exp. 10+ Years | =1 if the instructor has more than ten years of experience. | 0.42 | 0.49 |
| Age | The instructor's age in years. | 43.41 | 10.59 |

| | **Table 4 Continued** | | |
|---|---|---|---|
| Variable | Description | Mean | Std. Dev. |
| *Characteristics of the Riskiest Class Taught by the Instructor* | | | |
| F2F | =1 if the class is delivered in person. | 0.84 | 0.37 |
| Mixed | =1 if the class is delivered with a mix of in-person and online modalities. | 0.06 | 0.24 |
| Unspecified Modality | =1 if the class modality is not specified. | 0.10 | 0.30 |
| Safe Class | =1 if the ratio of No. Students to Safe Class Capacity is less than or equal to one. | 0.54 | 0.50 |
| Risky Class | =1 if the ratio of No. Students to Safe Class Capacity is greater than one. | 0.46 | 0.50 |
| Very Risky Class | =1 if the ratio of No. Students to Safe Class Capacity is greater than or equal to two. | 0.21 | 0.41 |
| No. Students | The number of students registered for the class | 28.29 | 37.74 |
| Safe Class Capacity | The number of students that can be seated in the classroom with six-foot social distancing | 20.63 | 12.76 |
| Classroom Area | The area of the classroom in thousands of square feet | 1.40 | 0.76 |
| Dispersible Class | =1 if the classroom furniture allows students to spread apart. | 0.25 | 0.43 |

The unit of observation is an instructor. The number of observations is 1,501.

**Table 5**
**Determinants of the Probability of Teaching a Risky or Very Risky Class in Spring 2021**

|  | (1) Risky | (2) Very Risky |
|---|---|---|
| *Individual Characteristics* | | |
| GTA | 0.066** | 0.063** |
|  | (0.032) | (0.029) |
| Adjunct Instructor | 0.085*** | 0.087*** |
|  | (0.028) | (0.027) |
| Lecturer | 0.009 | 0.036 |
|  | (0.028) | (0.024) |
| Staff | -0.066 | 0.045 |
|  | (0.060) | (0.055) |
| Female | 0.053** | 0.045* |
|  | (0.025) | (0.024) |
| Non-White | -0.008 | 0.006 |
|  | (0.019) | (0.017) |
| Age | -0.003*** | -0.002** |
|  | (0.001) | (0.001) |
| Exp. Less Than 1 Year | -0.019 | -0.044* |
|  | (0.030) | (0.025) |
| Exp. 1-3 Years | -0.001 | -0.005 |
|  | (0.024) | (0.020) |
| Exp. 3-5 Years | -0.017 | -0.042** |
|  | (0.024) | (0.020) |
| Exp. 5-10 Years | -0.014 | -0.024 |
|  | (0.025) | (0.021) |
| *Class Characteristics* | | |
| 1000-Level Class | 0.171*** | 0.086** |
|  | (0.044) | (0.037) |
| 2000-Level Class | 0.247*** | 0.161*** |
|  | (0.039) | (0.034) |
| 3000-Level Class | 0.158*** | 0.071*** |
|  | (0.028) | (0.019) |
| 4000-Level Class | 0.135*** | 0.045** |
|  | (0.028) | (0.021) |
| Master's Level Class | -0.050** | -0.043*** |
|  | (0.024) | (0.015) |
| F2F Modality | 0.493*** | 0.216*** |
|  | (0.022) | (0.017) |
| Mixed Modality | 0.569*** | 0.265*** |
|  | (0.039) | (0.035) |

|  | (1) | (2) |
|---|---|---|
|  | Risky | Very Risky |
| Unspecified Modality | 0.294*** | 0.137*** |
|  | (0.054) | (0.043) |
| N | 4,382 | 4,382 |
| Course Type FEs | Yes | Yes |
| Department FEs | Yes | Yes |

**Table 5 Continued**

The unit of observation is a class. All classes whose risk can be determined are included in the regressions. The outcome in column 1 (2) is an indicator that is equal to one if the actual number of students enrolled in the class is greater than (twice) the safe capacity, i.e., the maximum number of students that can be seated while maintaining a six-foot distance. Standard errors that are clustered at the instructor level are presented in parentheses. ***, **, and * indicate the statistical significance at 1%, 5%, and 10% levels, respectively.

**Table 6**
**The Probability of Teaching a Hypothetically Risky or Very Risky Class in Spring 2020**

|  | (1) Hypothetically Risky | (2) Hypothetically Very Risky |
|---|---|---|
| *Individual Characteristics* | | |
| GTA | 0.041 | -0.022 |
|  | (0.031) | (0.040) |
| Adjunct Instructor | 0.035 | -0.000 |
|  | (0.028) | (0.031) |
| Lecturer | 0.022 | 0.054* |
|  | (0.031) | (0.031) |
| Staff | 0.051 | 0.076 |
|  | (0.038) | (0.055) |
| Female | 0.007 | 0.002 |
|  | (0.017) | (0.019) |
| Non-White | 0.001 | -0.008 |
|  | (0.022) | (0.022) |
| Age | -0.001 | -0.001 |
|  | (0.001) | (0.001) |
| Exp. Less Than 1 Year | 0.035 | 0.019 |
|  | (0.033) | (0.033) |
| Exp. 1-3 Years | 0.054** | 0.017 |
|  | (0.026) | (0.027) |
| Exp. 3-5 Years | 0.003 | -0.041 |
|  | (0.025) | (0.027) |
| Exp. 5-10 Years | 0.033 | -0.007 |
|  | (0.025) | (0.028) |
| N | 4,055 | 4,055 |
| Other Control Variables | Yes | Yes |
| Course Type FEs | Yes | Yes |
| Department FEs | Yes | Yes |

The unit of observation is a class. All classes from Spring 2020 whose hypothetical risk can be determined are included in the regressions. The outcome in column 1 (2) is an indicator that is equal to one if the actual number of students enrolled in the class is greater than (twice) the safe capacity, i.e., the maximum number of students that can be seated while maintaining a six-foot distance. Standard errors that are clustered at the instructor level are presented in parentheses. ***, **, and * indicate the statistical significance at 1%, 5%, and 10% levels, respectively.

**Table 7**
**First Stage Regressions**

|  | (1) | (2) |
|---|---|---|
|  | Risky Class | Very Risky Class |
| Dispersible Class | -0.198*** | -0.152*** |
|  | (0.034) | (0.028) |
| N | 1,501 | 1,501 |
| F-Statistic | 34.94 | 28.50 |
| Control Variables | Yes | Yes |
| Course Type F.E.s | Yes | Yes |
| Department F.E.s | Yes | Yes |

The unit of observation is an instructor. The outcome in column 1 (2) is an indicator that is equal to one if the actual number of students enrolled in the riskiest class taught by the instructor is greater than (twice) the safe capacity, i.e., the maximum number of students that can be seated while maintaining a six-foot distance. *Dispersible Class* is an indicator that takes the value of one if students registered in the instructor riskiest class can spread away from each other when attending the lecture. Controls include the instructor's rank, sex, race, age, experience, the number of undergraduate and graduate classes taught by the instructor, the log of monthly wages earned by the instructor during the previous year, class characteristics, and the complete set of interactions between rank, age, and experience. Class characteristics (level, modality, course type, number of students, and classroom area) pertain to the riskiest class taught by the instructor. Refer to Appendix B Table B3 for the complete set of results. The F-Statistic presented is for the statistical significance of *Dispersible Class*. Robust standard errors are presented in parentheses. ***, **, and * indicate the statistical significance at 1%, 5%, and 10% levels, respectively.

**Table 8**
**2SLS Estimates of the Relationship between Classroom Risk and Instructor Pay**

|  | (1) | (2) | (3) |
|---|---|---|---|
|  | Second Stage | Second Stage | Reduced Form |
|  | Log(Wage) | Log(Wage) | Log(Wage) |
| Risky Class | 0.175** <br> (0.083) |  |  |
| Very Risky Class |  | 0.228** <br> (0.108) |  |
| Dispersible Class |  |  | -0.035** <br> (0.017) |
| N | 1,501 | 1,501 | 1,501 |
| Control Variables | Yes | Yes | Yes |
| Course Type FEs | Yes | Yes | Yes |
| Department FEs | Yes | Yes | Yes |

The unit of observation is an instructor. The outcomes are the logarithm of the monthly wages of the instructor. *Risky Class* (*Very Risky Class*) is an indicator that is equal to one if the actual number of students enrolled in the riskiest class taught by the instructor is greater than (twice) the safe capacity, i.e., the maximum number of students that can be seated while maintaining a six-foot distance. *Dispersible Class* is an indicator that takes the value of one if students registered in the instructor riskiest class can spread away from each other when attending the lecture. Control variables are the same as those in Table 7. Refer to Appendix B Table B5 for the complete set of results. Robust standard errors are presented in parentheses. ***, **, and * indicate the statistical significance at 1%, 5%, and 10% levels, respectively.

**Table 9**
**Relationship between (Hypothetical) Risk and Instructor Pay in Spring 2020**

|  | (1) | (2) |
|---|---|---|
|  | Log(Wage) | Log(Wage) |
| Hypothetically Risky Class | -0.146 |  |
|  | (0.127) |  |
| Hypothetically Very Risky Class |  | -0.152 |
|  |  | (0.133) |
| N | 1,618 | 1,618 |
| Control Variables | Yes | Yes |
| Course Type FEs | Yes | Yes |
| Department FEs | Yes | Yes |

The unit of observation is an instructor. Observations from Spring 2020 enter into regression. The outcomes are the logarithm of the monthly wages of the instructor. *Hypothetically Risky Class* (*Very Risky Class*) is an indicator that is equal to one if the actual number of students enrolled in the riskiest class taught by the instructor is greater than (twice) the safe capacity, i.e., the maximum number of students that can be seated while maintaining a six-foot distance. Results are from 2SLS using *Dispersible Class* as the instrument. Control variables are the same as those in Table 7. Refer to Appendix B Table B6 for the complete set of results. Robust standard errors are presented in parentheses. \*\*\*, \*\*, and \* indicate the statistical significance at 1%, 5%, and 10% levels, respectively.

# APPENDIX A

# Details on Auburn University

## *General Information*

Auburn University is a public (Carnegie R1) research university located in Auburn, Alabama. Established in 1856, the university comprises 206 academic buildings on 1,841 acres and has a student body of about 30,000. University employs 5,000 full-time workers, of which administrative/professional personnel make up about half, faculty personnel make up about a quarter, and staff personnel the remaining quarter.[56]

Operations of Auburn University are extremely important for the well-being of the local economy since the businesses around Auburn depend heavily on student demand. The ratio of the number of students to the population in the Auburn Metro Area, according to the 2020 Census, is about 20%. In addition, Auburn's home games in intercollegiate competitions attract tens of thousands of visitors to the town, generating positive externalities for the local food services and accommodations industries (Lentz and Laband 2009). According to an internal report, Auburn University's direct economic impact is $2.2 billion, "representing Auburn's in-state expenditures, such as payroll and purchases, student spending on local housing and food, construction, and spending by visitors to university events."[57]

## *The Shutdown in Spring 2020 and the Pressures to Reopen in the 2020-2021 Academic Year*

During the Spring 2020 semester, the university ceased on-campus instruction following Spring Break, as the state of Alabama entered a mandatory lockdown period.[58] The closure of the university had immediate and significant consequences, not only on students and faculty but also on local residents and businesses. The demographic makeup of the city in which the university is located displays a heavy concentration of college students relative to its population, compared to other US college towns. As such, the local economy was placed in a precarious situation while the campus was closed to students, many of whom left the area.[59]

---

[56] The details of these statistics can be found in this link.
[57] The announcement of the results of this study is in this link.
[58] See the Governor's Stay at Home Order.
[59] Business Insider ranked the city of Auburn as the 22nd most likely college town to be in financial ruin if it did not reopen campus in the Fall 2020 semester, given its high share of undergraduate students to population (36.4%). Similar findings are reported in this study, which ranked Auburn the 10th most vulnerable college town during the COVID-19 pandemic.

In Appendix A Figure A1, using Google's foot traffic data, we present the mobility trends in retail and recreation venues (as a proxy for local business demand) in Spring 2020 in the counties where AL colleges are located.[60] The vertical axis measures the percent change in mobility in the county of each university relative to the average in the first five weeks of 2020 in that same county. To eliminate daily fluctuations, we present the data as the average of the last seven days. The bold, dark-green line portrays Lee county, which is the home to Auburn University. The average of the whole state of Alabama is represented by the thin dashed line. The thin solid lines in various colors are the trends of the counties of other universities in Alabama.[61] Appendix A Figure A1 shows that all college counties, including Auburn University's, moved along the same pattern as the state average in Spring 2020. However, it is important to highlight that across all of these college towns, Auburn University's location, Lee county, experienced *the worst* decreases in magnitude in demand. In other words, the reduction in recreation and retail shops' foot traffic is the highest around Auburn University. This finding suggests that the local businesses were hit especially hard around Auburn, perhaps harder than others in the rest of the state.

Besides the local businesses, Auburn University's finances suffered during the shutdown in Spring 2020. For example, the top financial executive of the university reported that the university's losses amounted to $15 million through the mid-summer of 2020.[62] In addition, Auburn University was under pressure from parents and students who questioned the effectiveness of remote instruction.[63]

### *The Reopening of the University*

During the reopening, the university initiated a number of policies. When planning their courses for the Fall 2020 semester, instructors were allowed some leeway. They were asked to choose the modality they thought best aligned with their courses' objectives. However, this *laissez-faire* approach changed in Spring 2021 when the university required all instructors to teach their classes in pre-pandemic modalities (typically F2F) unless they have an exceptional medical excuse or a compelling pedagogical counter reason.[64] The consequences of this policy change are reflected in Table 1 in the main text, which displays the distribution of modalities by semester in the academic year 2020-2021. Specifically, the push for in-person classes by the university led the proportion of classes taught in-person to increase drastically, from 19% in Fall 2020 to 71% in Spring 2021.

---

[60] The data can be acquired from this link.
[61] These universities are the University of North Alabama, Alabama State University, Troy University, and the University of Alabama.
[62] The interview with the university official is at this link.
[63] Recent studies highlight the superiority of in-person over remote instruction (Altindag, Filiz and Tekin 2021; Kofoed et al. 2021; Bird et al. 2020).

Presumably to comply with the AL Governor's health order, that requires colleges to implement six-foot social distancing in the classroom, and at the same time to offer as many in-person classes as possible, Auburn University started measuring the safe capacities of each classroom in Summer 2020. At the June 16 University Senate meeting, the Provost announced that classrooms would practice safe social distancing and that a team was working on classroom diagrams.[65] Shortly afterward, the first CCAs were conducted by the Office of the University Architect. The goal of the surveys was to determine their safe capacities, i.e., the maximum number of students who can be seated in the classrooms while maintaining CDC-recommended six-foot social distancing. Examples of these studies are presented in Appendix A Figures A2 and A3.

Each room study has a timestamp indicating the day on which it was completed. Appendix A Figure A4 displays the timeline of these studies. According to these dates, the CCAs appear to have stopped on August 11, before the safe capacities of all classrooms were determined and the Fall 2020 semester began.[66] The studies were restarted on November 18, and the complete set was made available on the university's web page on December 19, long after classes of the Spring 2021 semester were scheduled. The university did not announce the completion of these studies to the campus community. Some members of the Auburn University community speculate that the university administration was trying to suppress this information from the instructors and students.

Despite the fact that the university administration had the information about the safe capacities of the classrooms, Auburn University did not implement the CDC-prescribed six feet social distancing order of the AL Governor in the 2020-2021 academic year. This is likely because of the capacity constraints of the university. For example, when asked about physical distancing in the classrooms at a General Faculty meeting on October 27, Auburn's Director of Facilities summarized the university's rationale: "In a classroom with fixed seating, the 6 feet *recommendation* is difficult to handle and still meet capacity needs [emphasis added]."[67]

---

[64] According to this policy, requests of modality change were approved or rejected by the deans of each college. Instructors with medical conditions had to explain them to the HR managers, and those with pedagogical reasons made their cases with their department chairs. Despite the fact that there has not been one single instructor whose modality change request was rejected, the policy certainly increased the cost of delivering classes in an online modality.
[65] These considerations were discussed in the University Senate's meeting on June 16, 2020.
[66] We found that 105 out of the 114 classrooms for which a study was available at the beginning of the Fall 2020 semester hosted a class size above its safe capacity.

In the Fall 2020 semester, the university initiated its "50% Rule" policy, which allowed classrooms to contain up to half of their usual enrollment limits.[68] According to this rule, a class will not be allowed to contain more than 50% of its usual student capacity. It is important to note that the 50% Rule was not sufficient to keep the number of students in a class below the safe capacity. For example, we found that 105 out of the 114 classrooms for which a study was available at the beginning of the Fall 2020 semester hosted a class size above its safe capacity. Despite this, the university administration did not change the 50% Rule for this academic year. In the Spring 2021 semester, the 50% rule was not implemented universally. We have identified at least 435 violating classes in that semester.

University's disregard for social distancing guidelines caused discontent among the faculty. For example, the Auburn chapter of the American Association of University Professors (AAUP) recommended that the university should adopt a number of policies regarding safety from COVID-19. In a special faculty meeting on November 10, 2020, several faculty expressed concerns about classroom capacities. In addition, this meeting generated a motion of "No Confidence" in the Provost.[69] As a response, the university administration held a series of Town Hall Meetings where the Provost and other members of the administration fielded questions from instructors.[70] In these meetings, Auburn University's central administrators announced that the enrollment decisions in the classrooms would be "pushed down" to the lowest level—i.e., the department and instructor level. In the first days of the Spring 2021 semester, during a University Senate meeting, the Director of Auburn's Medical Clinic and member of the COVID-19 task force explained that Auburn University does not have a policy with respect to social distancing in the classroom and informed faculty that they did not have the authority to enforce social distancing in their classrooms if students wanted to sit together.

---

[67] The director also stated in the same speech that "movable furniture is easier to make work" providing us with additional justification for our instrumental variable.

[68] By contrast, Jordan-Hare Stadium, an open-air facility on campus, was restricted to 20% capacity during the 2020 college football season.

[69] The motion of no confidence in provost was shut down by the chair of the senate in the meeting since the Provost was not attending that meeting. Faculty who were adamant about the motion petitioned for another special called meeting to debate and vote on the motion on November 17. Note that the CCAs resumed on November 18 after a pause since August 11.

[70] For video of two of those meetings, seethe Provost's Town Hall Dec. 3: https://archive.org/details/zoom_0_20201203_2319 and the Provost Town Hall Jan 27 https://archive.org/details/zoom_0_20210127_2238.

It is apparent that Auburn University's position on social distancing in the classroom evolved substantially between June 2020 and the preparations for the Spring 2021 semester. However, it is important to note that the AL Governor's "Safer at Home Order" maintains the same language, with respect to social distancing in the classroom, from the initial order in May 2020 until the latest extension of that order in January 2021 (which was in effect until March 2021). For the purpose of our study, the discretion that is given to faculty in choosing how they deal with social distancing and, particularly, the scheduling flexibility afforded department heads/chairs, likely creates heterogeneity in the assignment of risk across and within different administrative units at the university.

**Additional References for Appendix A**

Altindag, Duha Tore, Elif S. Filiz, and Erdal Tekin. (2021). "Is Online Education Working?" NBER Working Paper Series #29113.

Bird, Kelli A., Benjamin L. Castleman, and Gabrielle Lohner. (2020). "Negative Impacts From the Shift to Online Learning During the COVID-19 Crisis: Evidence from a Statewide Community College System." EdWorkingPaper No. 20-299. Annenberg Institute for School Reform at Brown University.

Kofoed, Michael S., Lucas Gebhart, Dallas Gilmore, and Ryan Moschitto. (2021). "Zooming to Class?: Experimental Evidence on College Students' Online Learning during COVID-19." IZA Discussion Papers 14356.

Lentz, Bernard F., and David N. Laband. (2009). "The Impact of Intercollegiate Athletics on Employment in the Restaurant and Accommodations Industries." *Journal of Sports Economics* 10(4), pp. 351-368.

**Appendix A Figure A1**
**Foot Traffic Trends in Recreation and Retail Venues in Alabama College Communities in Spring 2020**

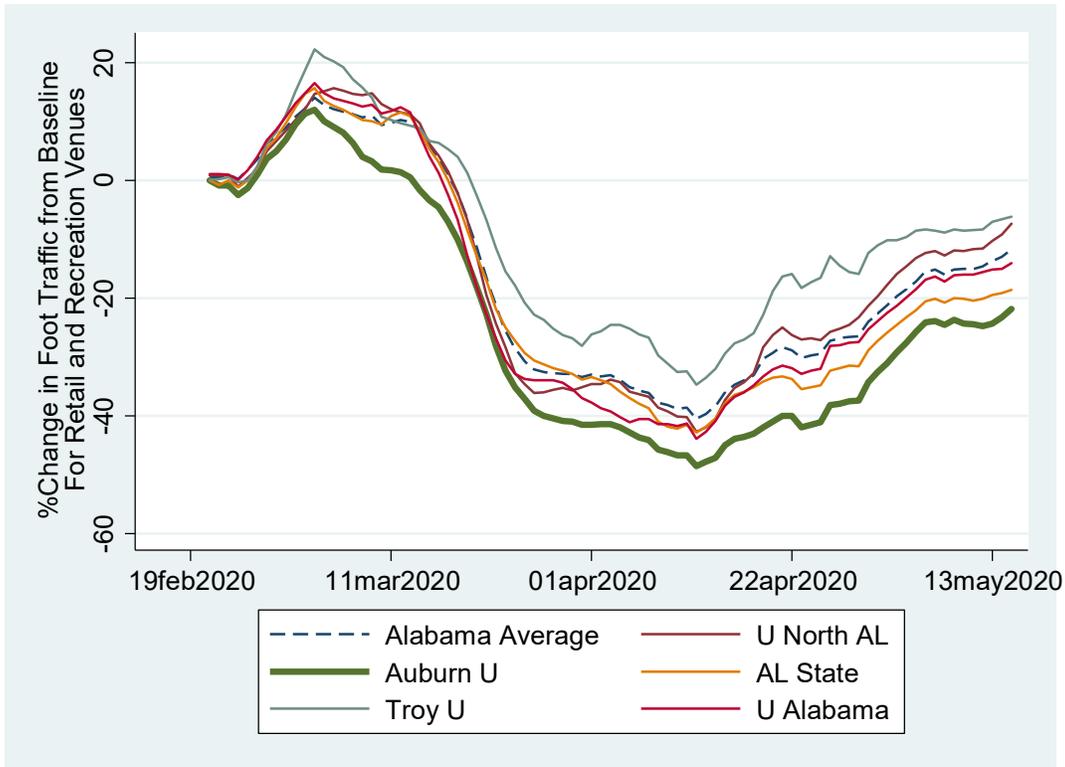

The source of the data is Google's Community Mobility Reports. We present the average of the last seven days. The vertical axis measures the change in foot traffic in retail and recreational venues relative to the baseline period (the first five weeks of 2020). The dashed line is the state average of Alabama. Solid lines represent the trends in the counties of the universities. The dark-green line displays the movements in Lee county, which is the home of Auburn University.

**Appendix A Figure A2**
**Example Classroom Capacity Analysis – Dispersible Classroom**

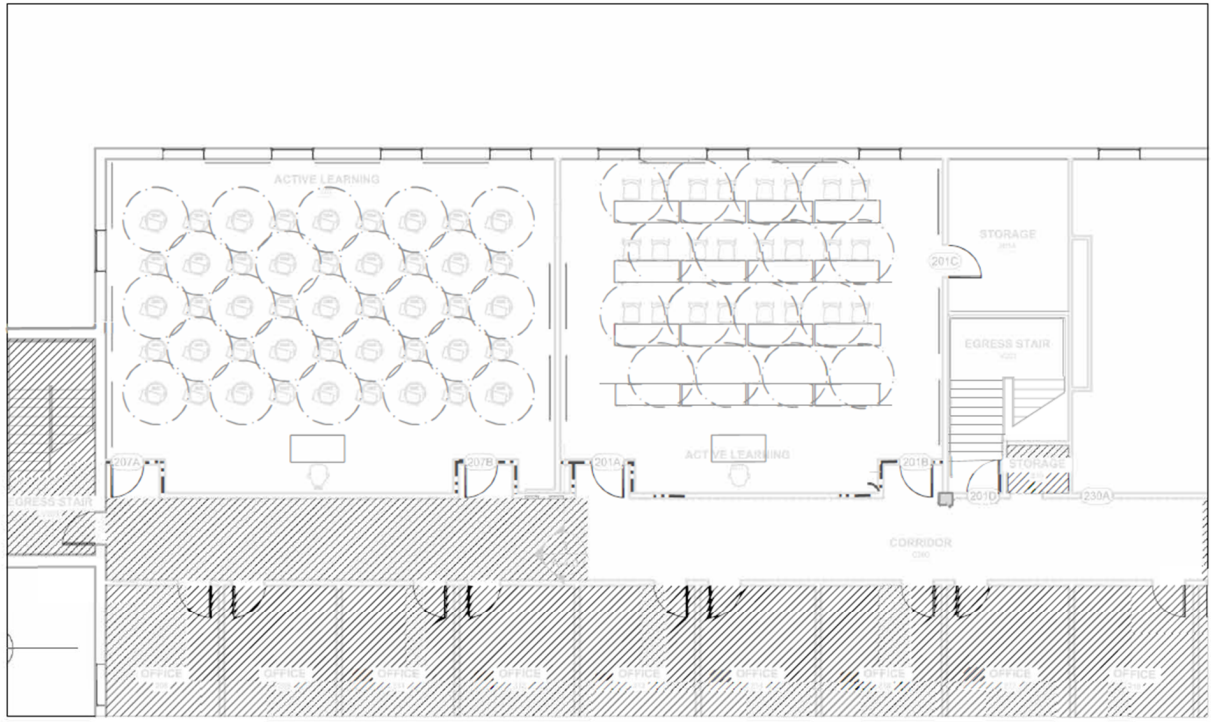

Classroom Capacity Analysis for Miller Hall 207. Source: Office of the University Architect at Auburn University. Accessed via this link.

**Appendix A Figure A3**
**Example Classroom Capacity Analysis – Non-Dispersible Classroom**

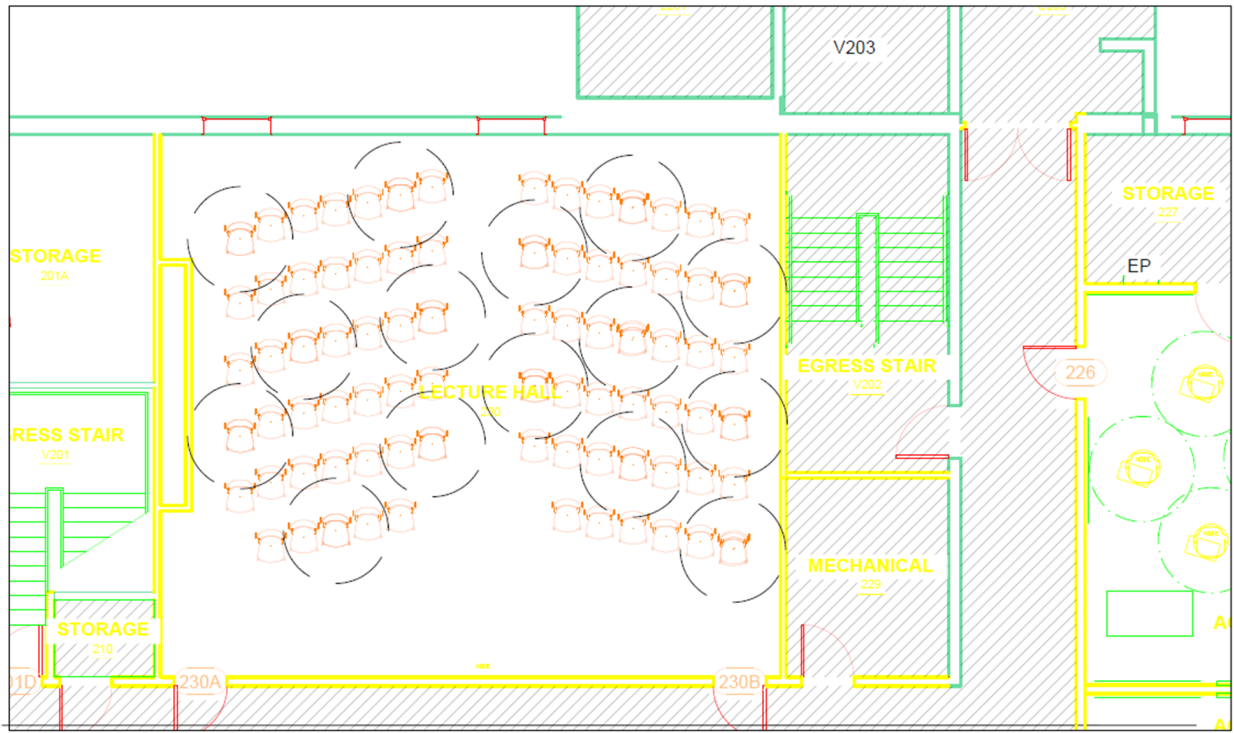

Classroom Capacity Analysis for Miller Hall 230. Source: Office of the University Architect at Auburn University. Accessed via this link.

**Appendix A Figure A4**

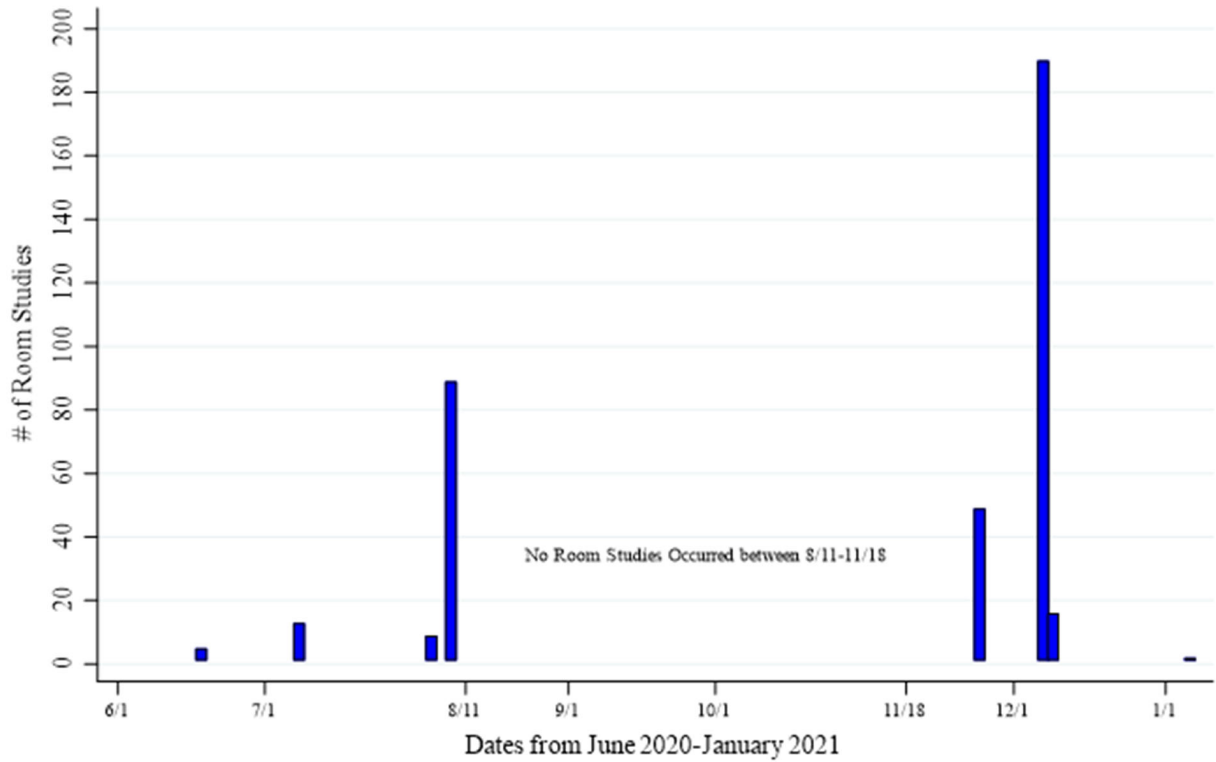

The blue bars tally the number of Classroom Capacity Analyses which were conducted by the Office of the University Architect each week between June 2020 and January 2021. All studies were completed as of January 6, 2021.

**Appendix A Figure A5**
**Fixed Furniture Classroom (Miller Hall 230)**

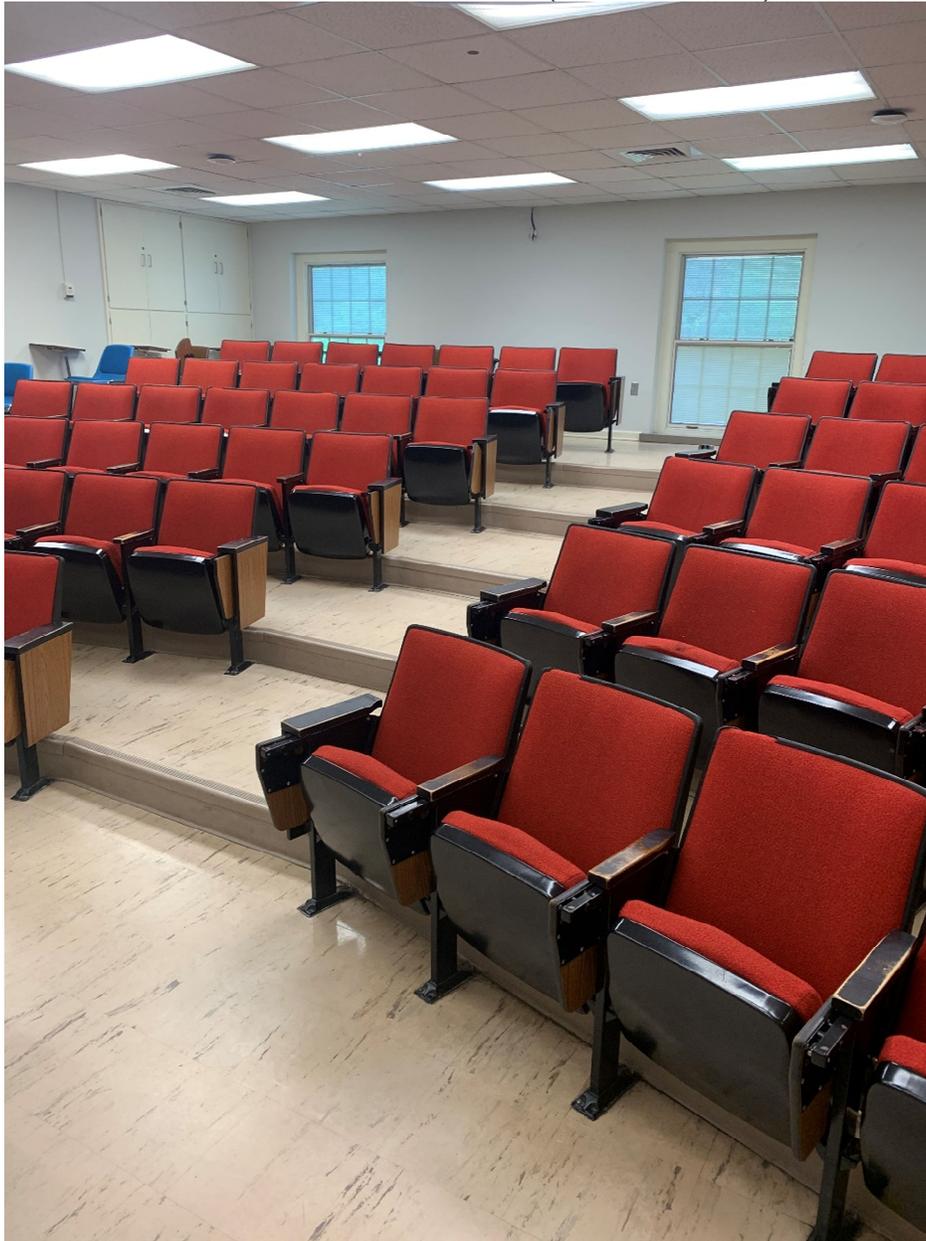

Room 230 in Miller Hall, Auburn University. Photographed by the authors on May 19, 2021.

**Appendix A Figure A6**
**Movable Furniture Classroom (Miller Hall 207)**

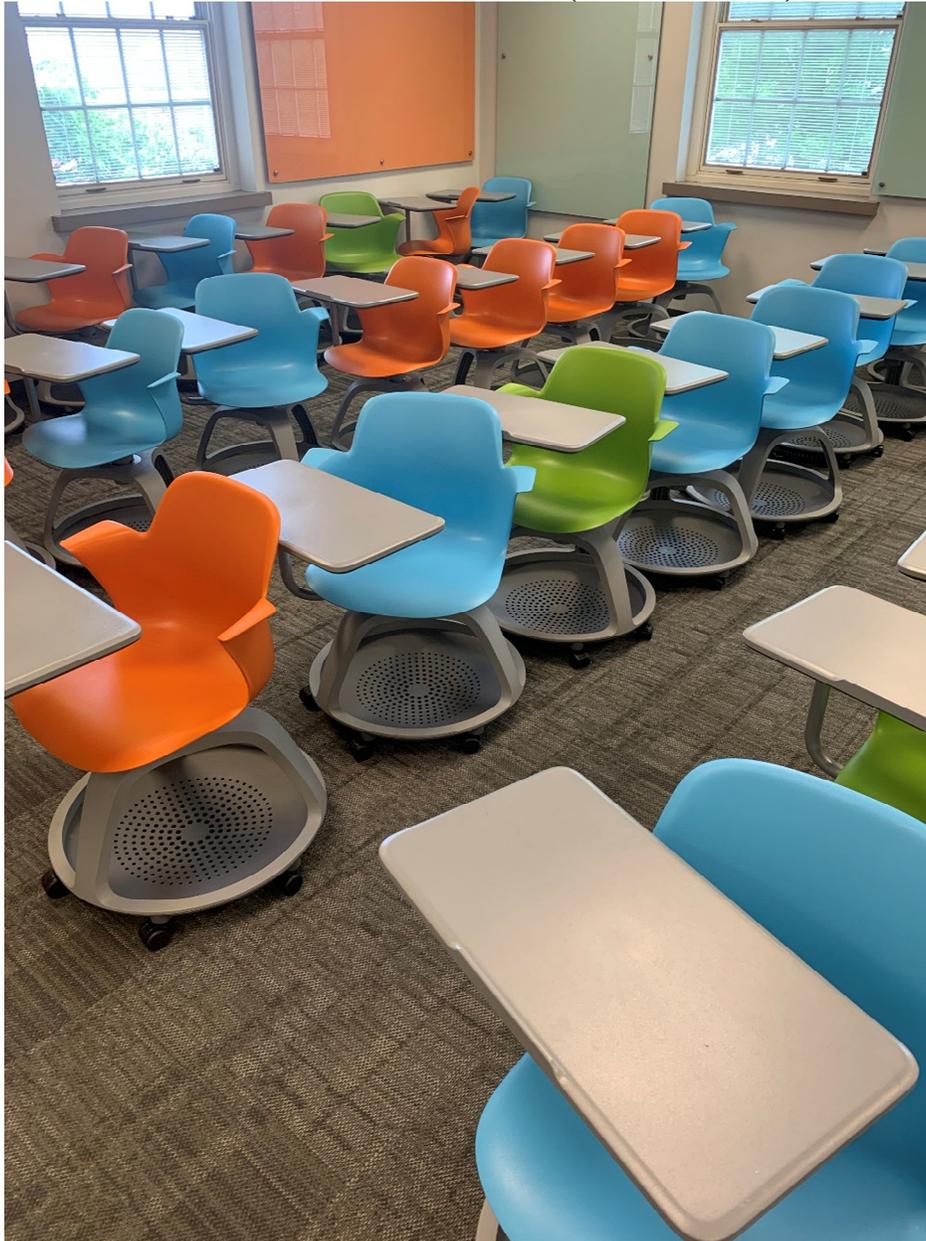

Room 207 in Miller Hall, Auburn University. Photographed by the authors on May 19, 2021.

## Appendix A Table A1
### Lecture Types

| Course Type | Number of Students |
|---|---|
| Lecture | 74,777 |
| Distance | 17,597 |
| Combined Lecture and Lab | 9,597 |
| Laboratory | 8,895 |
| Seminar | 1,019 |
| Studio | 842 |
| Special Topics | 635 |
| Practicum | 326 |
| Combined Lecture and Studio | 265 |
| Auburn Abroad Section | 124 |
| Internship | 122 |
| Independent Study | 117 |
| Dissertation | 43 |
| Clinic | 26 |
| Private Lesson | 16 |
| Field Experience | 13 |
| Masters Thesis | 2 |
| Qualifying Test | 0 |

Column 1 lists the 18 lecture-type categories for the Spring 2021 semester. Column 2 displays the number of students registered for each lecture type.

## Appendix A Table A2
## Subject Details

| Department/Subject | Percent of Classes that are Risky | Number of Students |
|---|---|---|
| Accounting | 55% | 3873 |
| Adult Education | 18% | 184 |
| Aerospace Engineering | 21% | 781 |
| Africana Studies | 0% | 13 |
| Agric Economics | 57% | 386 |
| Agriculture | 33% | 18 |
| Aerospace Studies (AFROTC) | 100% | 394 |
| Animal Sciences | 42% | 751 |
| Anthropology | 27% | 466 |
| Applied Biotechnology | 0% | 8 |
| Architecture | 20% | 644 |
| Interior Architecture | 0% | 26 |
| Art | 35% | 876 |
| Flight Education | 13% | 133 |
| Aviation Management | 74% | 998 |
| Bio & Ag Technology Management | 0% | 19 |
| Biochemistry | 67% | 481 |
| Biology | 22% | 9027 |
| Biomaterials and Packaging | 0% | 18 |
| Building Science | 70% | 1477 |
| Biosystems Engineering | 47% | 343 |
| Business Analytics | 61% | 2084 |
| Business Admin | 2% | 2709 |
| Consumer and Design Sciences | 42% | 1358 |
| Community and Civic Engagement | 100% | 17 |
| Chemistry | 73% | 4519 |
| Chemical Engineering | 36% | 897 |
| Civil Engineering | 19% | 1096 |
| Communication Disorders | 48% | 740 |
| Communication and Journalism | 70% | 413 |
| Communication | 73% | 2453 |
| Computer Sci & Software En | 21% | 3246 |
| Counselor Ed, Counseling Psych | 32% | 870 |
| Community Planning | 0% | 145 |
| Computer Science | 0% | 762 |
| Crop, Soil, and Environmental Sciences | 5% | 436 |
| Career and Technical | 23% | 901 |
| Early Childhood Educ | 19% | 150 |
| Elementary Education | 26% | 184 |
| English for Spkrs Other Lang | 0% | 35 |
| Middle School Educ | 0% | 5 |
| Music Education | 24% | 167 |

## Appendix A Table A2 Continued

| Department/Subject | Percent of Classes that are Risky | Number of Students |
|---|---|---|
| Reading Education | 27% | 105 |
| Secondary Education | 7% | 156 |
| Economics | 60% | 2726 |
| Educational Leadership | 0% | 157 |
| Educational Media | 8% | 129 |
| Interdepartmental Education | 0% | 128 |
| Electrical and Computer Eng. | 18% | 1486 |
| Entrepren. Family Business | 29% | 308 |
| English | 36% | 4916 |
| Engineering | 29% | 1561 |
| Entomology | 0% | 48 |
| Environmental Design | 64% | 306 |
| Environmental Science | 0% | 40 |
| Educ Psychology | 0% | 32 |
| Ed Res Methods & Analysis | 24% | 209 |
| Earth System Science | 33% | 26 |
| Exploratory | 0% | 85 |
| Food Science | 0% | 38 |
| Finance | 58% | 2765 |
| Fisheries, Aquaculture, and Aquatic Sciences | 20% | 133 |
| Foreign Lng-Arabic | 0% | 9 |
| Foreign Lng-Chinese | 0% | 112 |
| Foreign Lng-French | 0% | 209 |
| Foreign Lng-Global Cultures | 0% | 44 |
| Foreign Lng-German | 13% | 142 |
| Foreign Lng-Italian | 0% | 86 |
| Foreign Lng-Japanese | 0% | 112 |
| Foreign Lng-Korean | 17% | 70 |
| Foreign Lng-Latin | 0% | 43 |
| Foreign Lng-Russian | 0% | 15 |
| Foreign Lng-Spanish | 2% | 1074 |
| Forest Engineering | 0% | 27 |
| Forestry | 17% | 335 |
| Foundations of Educ. | 14% | 215 |
| Forestry & Wildlife Sci. | 0% | 214 |
| Graphic Design | 67% | 302 |
| Geography | 4% | 609 |
| Geology | 49% | 1798 |
| Auburn Global | 0% | 52 |
| Graduate Studies | 0% | 188 |
| Geospatial and Env Informatics | 0% | 14 |
| Global Studies/Human Sciences | 50% | 88 |
| Health Administration | 7% | 414 |

## Appendix A Table A2 Continued

| Department/Subject | Percent of Classes that are Risky | Number of Students |
|---|---|---|
| Human Dev & Family Studies | 34% | 1286 |
| Higher Education Admin | 33% | 134 |
| History | 36% | 3192 |
| Honor | 39% | 723 |
| Horticulture | 12% | 370 |
| Hospitality Management | 42% | 700 |
| Human Resource Mngt | 23% | 301 |
| Human Sciences, General | 0% | 98 |
| Interdisciplinary Studies | 0% | 55 |
| Industrial Design | 48% | 500 |
| Industrial & Sys Eng. | 19% | 1305 |
| Information Systems Management | 43% | 1715 |
| Journalism | 52% | 471 |
| Kinesiology | 7% | 3195 |
| Landscape Architecture | 36% | 84 |
| Liberal Arts, General | 92% | 344 |
| Laboratory Science | 50% | 50 |
| Leadership | 67% | 73 |
| Mathematics | 40% | 4064 |
| Materials Engineering | 14% | 354 |
| Media Studies | 71% | 486 |
| Mechanical Engineering | 22% | 1952 |
| Military Science (AROTC) | 91% | 185 |
| Marketing | 82% | 2419 |
| Management | 61% | 2013 |
| Music - Applied | 0% | 16 |
| Music Ensemble | 5% | 246 |
| Music | 26% | 1888 |
| Natural Resources Management | 18% | 261 |
| Naval Science (NROTC) | 71% | 237 |
| Nutrition | 24% | 571 |
| Nursing | 16% | 1423 |
| Polymer & Fiber Engineering | 25% | 11 |
| Physical Education | 3% | 1508 |
| Philosophy | 79% | 2468 |
| Physics | 3% | 1080 |
| Plant Pathology | 0% | 81 |
| Political Science | 43% | 1649 |
| Poultry Science | 17% | 91 |
| Public Relations Comm. | 64% | 517 |
| Psychology | 38% | 2577 |
| Interdept. Pharmacy | 0% | 56 |
| Pharmacy PharmD | 0% | 51 |

| Appendix A Table A2 Continued | | |
|---|---|---|
| Department/Subject | Percent of Classes that are Risky | Number of Students |
| Real Estate Development | 0% | 42 |
| Rehabilitation & Spec Educ | 11% | 774 |
| Rural Sociology | 0% | 29 |
| Sciences & Math | 0% | 1036 |
| Supply Chain Management | 68% | 1775 |
| Sociology | 29% | 655 |
| Social Work | 15% | 302 |
| Statistics | 17% | 1119 |
| Sustainability Studies | 20% | 111 |
| Theatre | 9% | 356 |
| University | 0% | 790 |
| VM | 0% | 33 |
| Veterinary Medicine | 0% | 481 |
| Wildlife Sciences | 29% | 398 |
| Women's Studies | 33% | 86 |

Column 1 lists the 143 Department/Subjects for the Spring 2021 semester. Column 2 displays their share of classes which are classified as *Risky*, i.e.,the actual number of students enrolled is greater than the safe capacity. Column 3 displays the number of students registered within each Department/Subject.

# APPENDIX B

## Additional Figures and Tables

### Appendix B Figure B1
### First Stage Estimates for Different Risk Thresholds

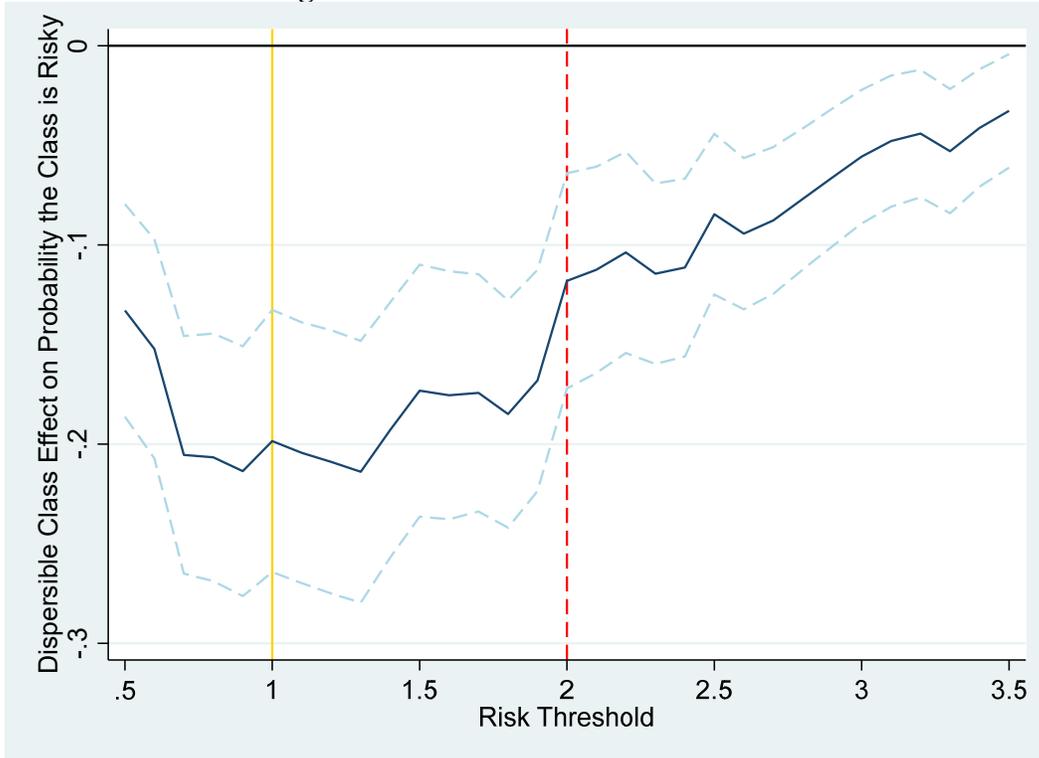

The solid blue line represents first stage point estimates for the relationship between classroom risk and dispersible classes. Dashed lines represent the 95% confidence intervals. The outcome, *Risky*, is equal to one if the ratio of the number of students in the class to the CDC safe capacity of the room exceeds the risk threshold denoted on the horizontal axis. All risk thresholds between 0.5 and 3.5 are considered in increments of 0.1.

**Appendix B Table B1**
**The Complete Set of Estimates in Table 6**
**(The Probability of Teaching a Hypothetically Risky or Very Risky Class in Spring 2020)**

|  | (1) Hypothetically Risky | (2) Hypothetically Very Risky |
|---|---|---|
| *Individual Characteristics* | | |
| GTA | 0.041 | -0.022 |
|  | (0.031) | (0.040) |
| Adjunct Instructor | 0.035 | -0.000 |
|  | (0.028) | (0.031) |
| Lecturer | 0.022 | 0.054* |
|  | (0.031) | (0.031) |
| Staff | 0.051 | 0.076 |
|  | (0.038) | (0.055) |
| Female | 0.007 | 0.002 |
|  | (0.017) | (0.019) |
| Non-White | 0.001 | -0.008 |
|  | (0.022) | (0.022) |
| Age | -0.001 | -0.001 |
|  | (0.001) | (0.001) |
| Exp. Less Than 1 Year | 0.035 | 0.019 |
|  | (0.033) | (0.033) |
| Exp. 1-3 Years | 0.054** | 0.017 |
|  | (0.026) | (0.027) |
| Exp. 3-5 Years | 0.003 | -0.041 |
|  | (0.025) | (0.027) |
| Exp. 5-10 Years | 0.033 | -0.007 |
|  | (0.025) | (0.028) |
| *Classroom Characteristics* | | |
| 1000-Level | 0.302*** | 0.264*** |
|  | (0.040) | (0.036) |
| 2000-Level | 0.254*** | 0.268*** |
|  | (0.041) | (0.036) |
| 3000-Level | 0.244*** | 0.173*** |
|  | (0.032) | (0.027) |
| 4000-Level | 0.216*** | 0.157*** |
|  | (0.031) | (0.028) |
| Master's Level | -0.030 | -0.028 |
|  | (0.029) | (0.022) |
| F2F Modality | 0.417*** | 0.217*** |
|  | (0.073) | (0.055) |
| N | 4,055 | 4,055 |
| Course Type FEs | Yes | Yes |
| Department FEs | Yes | Yes |

This table presents the complete set of estimates in Table 6. See notes to that table.

**Appendix B Table B2**
**OLS Estimates for the Relationship Between Classroom Risk and Instructor Pay**

|  | (1) | (2) |
|---|---|---|
|  | Log(Wage) | Log(Wage) |
| Risky Class | 0.011 |  |
|  | (0.019) |  |
| Very Risky Class |  | 0.025 |
|  |  | (0.017) |
| N | 1,501 | 1,501 |
| Other Control Variables | Yes | Yes |
| Course Type FEs | Yes | Yes |
| Department FEs | Yes | Yes |

The unit of observation is an instructor. The outcome is the log monthly wage of the instructor. Class characteristics pertain to the riskiest class taught by the instructor. Robust standard errors are presented in parentheses. ***, **, and * indicate the statistical significance at 1%, 5%, and 10% levels, respectively.

**Appendix B Table B3**
**The Estimates of the Complete Set of Variables in Table 7 (First Stage Regressions)**

| | (1) | (2) |
|---|---|---|
| | Risky Class | Very Risky Class |
| Dispersible Class | -0.198*** | -0.152*** |
| | (0.034) | (0.028) |
| GTA | -0.439*** | -0.383** |
| | (0.159) | (0.155) |
| Adjunct Instructor | 0.144 | 0.251* |
| | (0.123) | (0.143) |
| Lecturer | -0.042 | -0.090 |
| | (0.332) | (0.287) |
| Staff | -0.126 | 0.133 |
| | (0.227) | (0.204) |
| Female | 0.026 | 0.020 |
| | (0.020) | (0.019) |
| Non-White | 0.026 | 0.042** |
| | (0.023) | (0.021) |
| Age | -0.000 | 0.001 |
| | (0.001) | (0.001) |
| Exp. Less Than 1 Year | -0.301 | 0.109 |
| | (0.218) | (0.185) |
| Exp. 1-3 Years | 0.160 | 0.155 |
| | (0.105) | (0.096) |
| Exp. 3-5 Years | -0.103 | -0.110 |
| | (0.147) | (0.122) |
| Exp. 5-10 Years | -0.129 | 0.046 |
| | (0.106) | (0.097) |
| GTA x Exp. Less Than 1 Year | 0.212 | 0.125 |
| | (0.281) | (0.257) |
| GTA x Exp. 1-3 Years | 0.209 | 0.295 |
| | (0.204) | (0.185) |
| GTA x Exp. 3-5 Years | 0.521** | 0.784*** |
| | (0.242) | (0.221) |
| GTA x Exp. 5-10 Years | 0.698*** | 0.216 |
| | (0.209) | (0.213) |
| Adjunct Instructor x Exp. Less Than 1 Year | 0.518* | -0.498** |
| | (0.300) | (0.232) |
| Adjunct Instructor x Exp. 1-3 Years | -0.262* | -0.140 |
| | (0.159) | (0.164) |
| Adjunct Instructor x Exp. 3-5 Years | -0.128 | -0.136 |
| | (0.261) | (0.214) |
| Adjunct Instructor x Exp. 5-10 Years | -0.033 | 0.010 |
| | (0.191) | (0.188) |
| Lecturer x Exp. 1-3 Years | -0.036 | 0.338 |
| | (0.370) | (0.337) |

| | (1) | (2) |
|---|---|---|
| **Appendix B Table B3 Continued** | | |
| | Risky Class | Very Risky Class |
| Lecturer x Exp. 3-5 Years | -0.393 | 0.073 |
| | (0.429) | (0.344) |
| Lecturer x Exp. 5-10 Years | 0.185 | 0.190 |
| | (0.362) | (0.341) |
| Staff x Exp. Less Than 1 Year | 0.225* | 0.034 |
| | (0.126) | (0.113) |
| Staff x Exp. 1-3 Years | 0.161 | -0.061 |
| | (0.447) | (0.324) |
| Staff x Exp. 3-5 Years | 0.826 | 0.441 |
| | (0.532) | (0.682) |
| Staff x Exp. 5-10 Years | 0.201 | 0.098 |
| | (0.293) | (0.311) |
| GTA x Age | 0.019*** | 0.033*** |
| | (0.005) | (0.006) |
| Adjunct Instructor x Age | -0.001 | -0.003 |
| | (0.002) | (0.003) |
| Lecturer x Age | 0.000 | 0.003 |
| | (0.007) | (0.006) |
| Staff x Age | 0.003 | -0.002 |
| | (0.005) | (0.004) |
| Exp. Less Than 1 Year x Age | 0.001 | -0.002 |
| | (0.005) | (0.005) |
| Exp. 1-3 Years x Age | -0.004 | -0.003 |
| | (0.002) | (0.002) |
| Exp. 3-5 Years x Age | 0.002 | 0.002 |
| | (0.004) | (0.003) |
| Exp. 5-10 Years x Age | 0.003 | -0.001 |
| | (0.002) | (0.002) |
| GTA x Exp. Less Than 1 Year x Age | -0.004 | -0.035*** |
| | (0.011) | (0.008) |
| GTA x Exp. 1-3 Years x Age | -0.012* | -0.028*** |
| | (0.006) | (0.007) |
| GTA x Exp. 3-5 Years x Age | -0.022*** | -0.041*** |
| | (0.008) | (0.007) |
| GTA x Exp. 5-10 Years x Age | -0.026*** | -0.029*** |
| | (0.006) | (0.007) |
| Adjunct Instructor x Exp. Less Than 1 Year x Age | -0.008 | 0.009 |
| | (0.007) | (0.007) |
| Adjunct Instructor x Exp. 1-3 Years x Age | 0.003 | 0.003 |
| | (0.004) | (0.004) |
| Adjunct Instructor x Exp. 3-5 Years x Age | 0.003 | -0.000 |
| | (0.007) | (0.006) |

**Appendix B Table B3 Continued**

| | (1) Risky Class | (2) Very Risky Class |
|---|---|---|
| Adjunct Instructor x Exp. 5-10 Years x Age | 0.000 | -0.001 |
| | (0.005) | (0.005) |
| Lecturer x Exp. 1-3 Years x Age | 0.001 | -0.005 |
| | (0.008) | (0.008) |
| Lecturer x Exp. 3-5 Years x Age | 0.010 | -0.003 |
| | (0.010) | (0.008) |
| Lecturer x Exp. 5-10 Years x Age | -0.002 | -0.003 |
| | (0.008) | (0.008) |
| Staff x Exp. 1-3 Years x Age | -0.006 | -0.001 |
| | (0.012) | (0.010) |
| Staff x Exp. 3-5 Years x Age | -0.022 | -0.015 |
| | (0.013) | (0.019) |
| Staff x Exp. 5-10 Years x Age | -0.008 | -0.002 |
| | (0.007) | (0.007) |
| No. Undergrad. Classes | -0.007** | 0.001 |
| | (0.003) | (0.003) |
| No. Grad. Classes | -0.006 | 0.005 |
| | (0.005) | (0.004) |
| Lag of Log Wages | 0.023 | 0.013 |
| | (0.027) | (0.021) |
| 1000-Level | 0.411*** | 0.174*** |
| | (0.060) | (0.053) |
| 2000-Level | 0.304*** | 0.126*** |
| | (0.051) | (0.043) |
| 3000-Level | 0.271*** | 0.102*** |
| | (0.045) | (0.034) |
| 4000-Level | 0.244*** | 0.095*** |
| | (0.045) | (0.033) |
| Master's Level | 0.123** | -0.017 |
| | (0.051) | (0.032) |
| F2F Modality | 0.031 | 0.008 |
| | (0.046) | (0.036) |
| Mixed Modality | 0.048 | 0.082 |
| | (0.059) | (0.052) |
| No. Students | 0.045*** | 0.055*** |
| | (0.005) | (0.005) |
| Classroom Area (sq. feet) | -0.121*** | -0.150*** |
| | (0.018) | (0.019) |
| N | 1,501 | 1,501 |
| Department & Course Type FEs | Yes | Yes |

This table presents the complete set of estimates in Table 7. See notes to that table.

**Appendix B Table B4**
**The Estimates of the Complete Set of Variables in Figure 2**
**(Correlation Between the Instrument and Observables)**

|  | (1) | (2) | (3) |
|---|---|---|---|
|  | Administrators and Tenure Track Professors | Female | Non-White |
| Dispersible Class | -0.031 | -0.024 | 0.001 |
|  | (0.037) | (0.038) | (0.035) |
|  | [0.990] | [1.000] | [1.000] |
| Classroom Area | -0.109 | 0.003 | -0.015 |
|  | (0.021) | (0.019) | (0.018) |
| N | 1,501 | 1,501 | 1,501 |
| Department FEs | Yes | Yes | Yes |

|  | (4) | (5) | (6) |
|---|---|---|---|
|  | Age | Exp. Less Than 1 Year | Exp. 1-3 Years |
| Dispersible Class | -0.600 | 0.002 | 0.074 |
|  | (1.166) | (0.008) | (0.035) |
|  | [1.000] | [1.000] | [0.356] |
| Classroom Area | -0.486 | 0.004 | -0.012 |
|  | (0.662) | (0.004) | (0.015) |
| N | 1,501 | 1,501 | 1,501 |
| Department FEs | Yes | Yes | Yes |

|  | (7) | (8) | (9) |
|---|---|---|---|
|  | Exp. 3-5 Years | Exp. 5-10 Years | Exp. More Than 10 Years |
| Dispersible Class | -0.007 | -0.003 | -0.066 |
|  | (0.031) | (0.035) | (0.042) |
|  | [1.000] | [1.000] | [0.832] |
| Classroom Area | -0.005 | 0.013 | 0.000 |
|  | (0.015) | (0.018) | (0.020) |
| N | 1,501 | 1,501 | 1,501 |
| Department FEs | Yes | Yes | Yes |

| | (10) | (11) | (12) |
|---|---|---|---|
| | Lag of Log Wages | 1000-Level Class | 2000-Level Class |
| Dispersible Class | -0.074 | -0.018 | -0.001 |
| | (0.059) | (0.024) | (0.029) |
| | [0.931] | [0.990] | [1.000] |
| Classroom Area | -0.084 | 0.059 | 0.080 |
| | (0.029) | (0.015) | (0.017) |
| N | 1,501 | 1,501 | 1,501 |
| Department FEs | Yes | Yes | Yes |

| | (13) | (14) | (15) |
|---|---|---|---|
| | 3000-Level Class | 4000-Level Class | Master's Level Class |
| Dispersible Class | -0.041 | -0.029 | 0.007 |
| | (0.031) | (0.030) | (0.027) |
| | [0.930] | [0.990] | [1.000] |
| Classroom Area | 0.009 | -0.040 | -0.037 |
| | (0.016) | (0.010) | (0.010) |
| N | 1,501 | 1,501 | 1,501 |
| Department FEs | Yes | Yes | Yes |

| | (16) | (17) |
|---|---|---|
| | Number of Students | Number of Students Allowed to Register |
| Dispersible Class | 0.150 | 3.399 |
| | (0.206) | (2.243) |
| | [1.000] | [0.921] |
| Classroom Area | 2.907 | 32.215 |
| | (0.283) | (3.093) |
| N | 1,501 | 1,501 |
| Department FEs | Yes | Yes |

This table presents the complete set of estimates in Figure 2. Robust standard errors are in parentheses. P-values, adjusted for multiple hypothesis testing, are displayed in brackets.

## Appendix B Table B5
## The Estimates of the Complete Set of Variables in Table 8
## (2SLS Estimates of the Relationship between Classroom Risk and Instructor Pay)

|  | (1) | (2) | (3) |
|---|---|---|---|
|  | Log(Wage) | Log(Wage) | Log(Wage) |
| Risky Class | 0.175** | | |
|  | (0.083) | | |
| Very Risky Class | | 0.228** | |
|  | | (0.108) | |
| Dispersible Class | | | -0.035** |
|  | | | (0.017) |
| GTA | 0.186 | 0.197 | 0.110 |
|  | (0.147) | (0.152) | (0.150) |
| Adjunct Instructor | -0.220 | -0.252 | -0.195 |
|  | (0.159) | (0.169) | (0.178) |
| Lecturer | -0.243*** | -0.230** | -0.250*** |
|  | (0.094) | (0.101) | (0.094) |
| Staff | 0.001 | -0.051 | -0.021 |
|  | (0.110) | (0.118) | (0.107) |
| Female | -0.025* | -0.026* | -0.021 |
|  | (0.014) | (0.014) | (0.014) |
| Non-White | 0.003 | -0.002 | 0.008 |
|  | (0.015) | (0.016) | (0.015) |
| Age | -0.000 | -0.001 | -0.001 |
|  | (0.001) | (0.001) | (0.001) |
| Exp. Less Than 1 Year | 0.240 | 0.163 | 0.188 |
|  | (0.416) | (0.420) | (0.429) |
| Exp. 1-3 Years | -0.087 | -0.094 | -0.059 |
|  | (0.065) | (0.066) | (0.065) |
| Exp. 3-5 Years | -0.320* | -0.313* | -0.338* |
|  | (0.165) | (0.167) | (0.174) |
| Exp. 5-10 Years | 0.155* | 0.122 | 0.132 |
|  | (0.090) | (0.088) | (0.096) |
| GTA x Exp. Less Than 1 Year | -0.554 | -0.545 | -0.517 |
|  | (0.440) | (0.449) | (0.455) |
| GTA x Exp. 1-3 Years | -0.271* | -0.302** | -0.235 |
|  | (0.144) | (0.147) | (0.147) |
| GTA x Exp. 3-5 Years | -0.095 | -0.182 | -0.004 |
|  | (0.210) | (0.220) | (0.223) |
| GTA x Exp. 5-10 Years | -0.691*** | -0.619*** | -0.569*** |
|  | (0.162) | (0.158) | (0.162) |
| Adjunct Instructor x Exp. Less Than 1 Year | -0.065 | 0.139 | 0.026 |
|  | (0.506) | (0.494) | (0.516) |
| Adjunct Instructor x Exp. 1-3 Years | 0.215 | 0.201 | 0.169 |
|  | (0.197) | (0.205) | (0.216) |

| | (1) | (2) | (3) |
|---|---|---|---|
| | Log(Wage) | Log(Wage) | Log(Wage) |
| Adjunct Instructor x Exp. 3-5 Years | 0.501* | 0.509* | 0.478* |
| | (0.258) | (0.266) | (0.281) |
| Adjunct Instructor x Exp. 5-10 Years | -0.133 | -0.141 | -0.139 |
| | (0.220) | (0.228) | (0.249) |
| Lecturer x Exp. 1-3 Years | 0.349*** | 0.266** | 0.343*** |
| | (0.112) | (0.125) | (0.116) |
| Lecturer x Exp. 3-5 Years | -0.214 | -0.299 | -0.283 |
| | (0.265) | (0.269) | (0.295) |
| Lecturer x Exp. 5-10 Years | 0.014 | 0.003 | 0.046 |
| | (0.147) | (0.154) | (0.151) |
| Staff x Exp. Less Than 1 Year | 0.547*** | 0.579*** | 0.586*** |
| | (0.103) | (0.100) | (0.103) |
| Staff x Exp. 1-3 Years | -0.095 | -0.053 | -0.067 |
| | (0.349) | (0.406) | (0.417) |
| Staff x Exp. 3-5 Years | 0.383 | 0.427 | 0.527 |
| | (0.349) | (0.342) | (0.350) |
| Staff x Exp. 5-10 Years | -0.221 | -0.209 | -0.186 |
| | (0.162) | (0.169) | (0.164) |
| GTA x Age | -0.017*** | -0.021*** | -0.014*** |
| | (0.004) | (0.005) | (0.004) |
| Adjunct Instructor x Age | -0.002 | -0.001 | -0.002 |
| | (0.003) | (0.003) | (0.003) |
| Lecturer x Age | 0.004* | 0.003 | 0.004* |
| | (0.002) | (0.002) | (0.002) |
| Staff x Age | -0.001 | -0.000 | -0.001 |
| | (0.002) | (0.002) | (0.002) |
| Exp. Less Than 1 Year x Age | -0.011 | -0.010 | -0.011 |
| | (0.010) | (0.010) | (0.011) |
| Exp. 1-3 Years x Age | 0.002 | 0.002 | 0.001 |
| | (0.002) | (0.002) | (0.002) |
| Exp. 3-5 Years x Age | 0.007* | 0.007* | 0.007* |
| | (0.004) | (0.004) | (0.004) |
| Exp. 5-10 Years x Age | -0.004* | -0.003 | -0.003 |
| | (0.002) | (0.002) | (0.002) |
| GTA x Exp. Less Than 1 Year x Age | 0.030** | 0.038*** | 0.030** |
| | (0.013) | (0.013) | (0.013) |
| GTA x Exp. 1-3 Years x Age | 0.013*** | 0.017*** | 0.011** |
| | (0.004) | (0.005) | (0.005) |
| GTA x Exp. 3-5 Years x Age | 0.010 | 0.015** | 0.006 |
| | (0.006) | (0.007) | (0.007) |
| GTA x Exp. 5-10 Years x Age | 0.021*** | 0.023*** | 0.016*** |
| | (0.005) | (0.006) | (0.005) |

**Appendix B Table B5 Continued**

**Appendix B Table B5 Continued**

| | (1) | (2) | (3) |
|---|---|---|---|
| | Log(Wage) | Log(Wage) | Log(Wage) |
| Adjunct Instructor x Exp. Less Than 1 Year x Age | 0.006 | 0.002 | 0.004 |
| | (0.012) | (0.012) | (0.013) |
| Adjunct Instructor x Exp. 1-3 Years x Age | -0.001 | -0.002 | -0.001 |
| | (0.004) | (0.004) | (0.005) |
| Adjunct Instructor x Exp. 3-5 Years x Age | -0.007 | -0.007 | -0.007 |
| | (0.006) | (0.007) | (0.007) |
| Adjunct Instructor x Exp. 5-10 Years x Age | 0.010** | 0.010** | 0.010* |
| | (0.005) | (0.005) | (0.006) |
| Lecturer x Exp. 1-3 Years x Age | -0.009*** | -0.007** | -0.009*** |
| | (0.003) | (0.003) | (0.003) |
| Lecturer x Exp. 3-5 Years x Age | 0.003 | 0.006 | 0.005 |
| | (0.006) | (0.006) | (0.007) |
| Lecturer x Exp. 5-10 Years x Age | -0.002 | -0.002 | -0.003 |
| | (0.004) | (0.004) | (0.004) |
| Staff x Exp. 1-3 Years x Age | 0.003 | 0.002 | 0.002 |
| | (0.012) | (0.013) | (0.014) |
| Staff x Exp. 3-5 Years x Age | -0.010 | -0.010 | -0.014 |
| | (0.008) | (0.008) | (0.008) |
| Staff x Exp. 5-10 Years x Age | 0.005 | 0.004 | 0.003 |
| | (0.004) | (0.004) | (0.004) |
| No. Undergrad. Classes | 0.000 | -0.001 | -0.001 |
| | (0.001) | (0.001) | (0.001) |
| No. Grad. Classes | 0.004* | 0.002 | 0.003 |
| | (0.003) | (0.003) | (0.003) |
| Lag of Log Wages | 0.796*** | 0.797*** | 0.800*** |
| | (0.045) | (0.045) | (0.048) |
| 1000-Level | -0.009 | 0.023 | 0.063* |
| | (0.048) | (0.039) | (0.035) |
| 2000-Level | -0.067* | -0.043 | -0.014 |
| | (0.038) | (0.031) | (0.028) |
| 3000-Level | -0.031 | -0.007 | 0.017 |
| | (0.033) | (0.026) | (0.024) |
| 4000-Level | -0.012 | 0.009 | 0.031 |
| | (0.029) | (0.023) | (0.022) |
| Master's Level | -0.014 | 0.011 | 0.008 |
| | (0.023) | (0.021) | (0.021) |
| F2F Modality | -0.040* | -0.036* | -0.034 |
| | (0.021) | (0.021) | (0.022) |
| Mixed Modality | -0.029 | -0.040 | -0.021 |
| | (0.029) | (0.029) | (0.029) |
| No. Students | -0.005 | -0.010 | 0.003 |
| | (0.004) | (0.006) | (0.002) |

**Appendix B Table B5 Continued**

|  | (1) | (2) | (3) |
|---|---|---|---|
|  | Log(Wage) | Log(Wage) | Log(Wage) |
| Classroom Area (sq. feet) | 0.009 | 0.022 | -0.012 |
|  | (0.012) | (0.016) | (0.010) |
| N | 1,501 | 1,501 | 1,501 |
| Course Type FEs | Yes | Yes | Yes |
| Department FEs | Yes | Yes | Yes |

This table presents the complete set of estimates in Table 8. See notes to that table.

**Appendix B Table B6**
**The Complete Set of Estimates in Table 9**
**Relationship between (Hypothetical) Risk and Instructor Pay in Spring 2020**

| | (1) | (2) |
|---|---|---|
| | Log(Wage) | Log(Wage) |
| Hypothetically Risky Class | -0.146 | |
| | (0.127) | |
| Hypothetically Very Risky Class | | -0.152 |
| | | (0.133) |
| GTA | 0.190* | 0.147 |
| | (0.103) | (0.105) |
| Adjunct Instructor | -0.114 | -0.141 |
| | (0.111) | (0.101) |
| Lecturer | -0.138 | -0.110 |
| | (0.102) | (0.112) |
| Staff | 0.041 | 0.076 |
| | (0.117) | (0.125) |
| Female | -0.015 | -0.014 |
| | (0.012) | (0.013) |
| Non-White | -0.022 | -0.022 |
| | (0.014) | (0.015) |
| Age | -0.001 | -0.001 |
| | (0.001) | (0.001) |
| Exp. Less Than 1 Year | -0.232 | -0.221 |
| | (0.146) | (0.145) |
| Exp. 1-3 Years | -0.007 | -0.018 |
| | (0.081) | (0.079) |
| Exp. 3-5 Years | -0.043 | -0.038 |
| | (0.109) | (0.109) |
| Exp. 5-10 Years | 0.002 | 0.002 |
| | (0.128) | (0.131) |
| GTA x Exp. Less Than 1 Year | -0.181 | -0.154 |
| | (0.204) | (0.202) |
| GTA x Exp. 1-3 Years | -0.220* | -0.189 |
| | (0.130) | (0.134) |
| GTA x Exp. 3-5 Years | -0.443*** | -0.420*** |
| | (0.139) | (0.143) |
| GTA x Exp. 5-10 Years | -0.212 | -0.211 |
| | (0.166) | (0.162) |
| Adjunct Instructor x Exp. Less Than 1 Year | 0.365* | 0.372** |
| | (0.191) | (0.190) |
| Adjunct Instructor x Exp. 1-3 Years | 0.279 | 0.321 |
| | (0.201) | (0.205) |
| Adjunct Instructor x Exp. 3-5 Years | -0.100 | -0.145 |
| | (0.183) | (0.184) |

**Appendix B Table B6 Continued**

| | (1) Log(Wage) | (2) Log(Wage) |
|---|---|---|
| Adjunct Instructor x Exp. 5-10 Years | 0.330* | 0.343* |
| | (0.190) | (0.195) |
| Lecturer x Exp. Less Than 1 Year | 0.455** | 0.465** |
| | (0.199) | (0.201) |
| Lecturer x Exp. 1-3 Years | -0.485* | -0.490* |
| | (0.251) | (0.251) |
| Lecturer x Exp. 3-5 Years | 0.570*** | 0.478** |
| | (0.170) | (0.193) |
| Lecturer x Exp. 5-10 Years | 0.154 | 0.121 |
| | (0.158) | (0.166) |
| Staff x Exp. Less Than 1 Year | -1.110* | -1.088* |
| | (0.656) | (0.655) |
| Staff x Exp. 1-3 Years | 0.067 | 0.110 |
| | (0.307) | (0.293) |
| Staff x Exp. 3-5 Years | -0.362 | -0.288 |
| | (0.248) | (0.237) |
| Staff x Exp. 5-10 Years | -0.116 | -0.119 |
| | (0.189) | (0.187) |
| GTA x Age | -0.010** | -0.010** |
| | (0.005) | (0.005) |
| Adjunct Instructor x Age | -0.003 | -0.002 |
| | (0.002) | (0.002) |
| Lecturer x Age | 0.002 | 0.001 |
| | (0.002) | (0.003) |
| Staff x Age | -0.003 | -0.003 |
| | (0.002) | (0.003) |
| Exp. Less Than 1 Year x Age | 0.005 | 0.005 |
| | (0.003) | (0.003) |
| Exp. 1-3 Years x Age | -0.000 | -0.000 |
| | (0.002) | (0.002) |
| Exp. 3-5 Years x Age | 0.000 | 0.000 |
| | (0.003) | (0.003) |
| Exp. 5-10 Years x Age | -0.000 | -0.000 |
| | (0.003) | (0.003) |
| GTA x Exp. Less Than 1 Year x Age | 0.004 | 0.004 |
| | (0.007) | (0.007) |
| GTA x Exp. 1-3 Years x Age | 0.005 | 0.005 |
| | (0.005) | (0.005) |
| GTA x Exp. 3-5 Years x Age | 0.011** | 0.011* |
| | (0.006) | (0.006) |
| Adjunct Instructor x Exp. Less Than 1 Year x Age | -0.008* | -0.007* |
| | (0.004) | (0.004) |

**Appendix B Table B6 Continued**

| | (1) | (2) |
|---|---|---|
| | Log(Wage) | Log(Wage) |
| Adjunct Instructor x Exp. 1-3 Years x Age | -0.004 | -0.004 |
| | (0.005) | (0.005) |
| Adjunct Instructor x Exp. 3-5 Years x Age | 0.003 | 0.005 |
| | (0.005) | (0.005) |
| Adjunct Instructor x Exp. 5-10 Years x Age | -0.004 | -0.004 |
| | (0.005) | (0.005) |
| Lecturer x Exp. Less Than 1 Year x Age | -0.010** | -0.010* |
| | (0.005) | (0.005) |
| Lecturer x Exp. 1-3 Years x Age | 0.011* | 0.012** |
| | (0.006) | (0.006) |
| Lecturer x Exp. 3-5 Years x Age | -0.013*** | -0.010** |
| | (0.004) | (0.005) |
| Lecturer x Exp. 5-10 Years x Age | -0.004 | -0.003 |
| | (0.004) | (0.004) |
| Staff x Exp. Less Than 1 Year x Age | 0.035* | 0.034* |
| | (0.020) | (0.020) |
| Staff x Exp. 1-3 Years x Age | 0.002 | -0.000 |
| | (0.010) | (0.010) |
| Staff x Exp. 3-5 Years x Age | 0.010 | 0.008 |
| | (0.007) | (0.006) |
| Staff x Exp. 5-10 Years x Age | 0.004 | 0.003 |
| | (0.004) | (0.004) |
| No. Undergrad. Classes | -0.003 | -0.002 |
| | (0.002) | (0.002) |
| No. Grad. Classes | 0.004 | 0.004 |
| | (0.003) | (0.003) |
| Lag of Log Wages | 0.837*** | 0.839*** |
| | (0.029) | (0.029) |
| 1000-Level Class | 0.056 | 0.059 |
| | (0.042) | (0.044) |
| 2000-Level Class | 0.002 | 0.008 |
| | (0.035) | (0.039) |
| 3000-Level Class | 0.052 | 0.049 |
| | (0.032) | (0.031) |
| 4000-Level Class | 0.024 | 0.015 |
| | (0.032) | (0.029) |
| Master's Level Class | 0.007 | 0.003 |
| | (0.021) | (0.021) |
| No. Students | 0.004 | 0.007 |
| | (0.006) | (0.008) |

|  | (1) | (2) |
|---|---|---|
| **Appendix B Table B6 Continued** | | |
|  | Log(Wage) | Log(Wage) |
| Classroom Area (sq. feet) | -0.020 | -0.022 |
|  | (0.022) | (0.024) |
| N | 1,618 | 1,618 |
| Course Type FEs | Yes | Yes |
| Department FEs | Yes | Yes |

This table presents the complete set of estimates in Table 9. See notes to that table.

**Appendix B Table B7**
**OLS and 2SLS Estimates of the Relationship between Continuous Classroom Risk and Instructor Pay**

|  | (1) | (2) | (3) | (4) |
|---|---|---|---|---|
|  | OLS | First Stage | 2SLS | Reduced Form |
|  | Log(Wage) | Classroom Risk | Log(Wage) | Log(Wage) |
| Classroom Risk | 0.008 |  | 0.076** |  |
|  | (0.009) |  | (0.035) |  |
| Dispersible Class |  | -0.455*** |  | -0.035** |
|  |  | (0.048) |  | (0.017) |
| N | 1,501 | 1,501 | 1,501 | 1,501 |
| First Stage F-Statistic |  | 90.97 |  |  |
| Control Variables | Yes | Yes | Yes | Yes |
| Course Type FEs | Yes | Yes | Yes | Yes |
| Department FEs | Yes | Yes | Yes | Yes |

The unit of observation is an instructor. *Classroom Risk* is defined as the ratio of the number of students in the class to the CDC safe capacity of the room. OLS estimation results are reported in column 1. First stage estimates are reported in column 2. 2SLS estimates are reported in column 3. Reduced form estimates are reported in column 4. Control variables are the same as those in Table 7. Robust standard errors are presented in parentheses. ***, **, and * indicate the statistical significance at 1%, 5%, and 10% levels, respectively.